%% file: ms.tex
\documentclass[12pt,preprint]{aastex}
\usepackage{psfig}

\begin{document}
\title{~~\\ ~~\\ Star Formation in H{\sc i} Selected Galaxies I: Sample Characteristics}
\author{J. F. Helmboldt\altaffilmark{1} \& R. A. M. Walterbos\altaffilmark{1}}
\email{helmbold@nmsu.edu, rwalteb@nmsu.edu}
\author{G. D. Bothun\altaffilmark{2}}
\email{nuts@bigmoo.uoregon.edu}
\author{K. O'Neil\altaffilmark{3}}
\email{koneil@gb.nrao.edu}
\author{W. J. G. de Blok\altaffilmark{4}}
\email{Erwin.deBlok@astro.cf.ac.uk}

\altaffiltext{1}{Department of Astronomy, New Mexico State University, Dept. 4500 PO Box 30001, Las Cruces, NM 88003}
\altaffiltext{2}{Department of Physics, University of Oregon, Eugene, Oregon 97403}
\altaffiltext{3}{NRAO, P.O. Box 2 Green Bank, WV 24944 }
\altaffiltext{4}{Department of Physics and Astronomy, Cardiff University, 5 The Parade, Cardiff CF24 3YB, United Kingdom}

\received{?}
\accepted{?}


\begin{abstract}

A sample of 69 galaxies with radial velocities less than 2500 km s$^{-1}$ was selected from the H{\sc i} Parkes All Sky Survey (H{\sc i}PASS) and imaged in broad band B and R and narrow band H$\alpha$ to deduce details about star formation in nearby disk galaxies while avoiding surface brightness selection effects.  The sample is dominated by late-type, dwarf disks (mostly Sc and Sm galaxies) with exponential disk scale lengths $\sim$1 to 5 kpc.  The H{\sc i}PASS galaxies on average have lower star formation rates (SFRs) and are bluer and lower surface brightness than an optically selected sample.  H{\sc ii} regions were detected in all but one of the galaxies.  Many galaxies had as few as two to five H{\sc ii} regions.  The galaxies' H$\alpha$ equivalent widths, colors, and SFRs per unit H{\sc i} mass are best explained by young mean ages ($\sim$3 to 5 Gyr according to Schmidt Law models) with star formation histories where the SFRs were higher in the past.  Comparison of the surface brightness coverage of the H{\sc i}PASS galaxies with that of an optically selected sample shows that such a sample may miss $\sim$10\% of the local galaxy number density and could possibly miss as much as 3 to 4\% of the SFR density.  The amount lower surface brightness galaxies contribute to the total luminosity density may be insignificant, but this conclusion is somewhat dependent on how the fluxes of these objects are determined.

\end{abstract}


\section{Introduction}

Ground based optical observations of galaxies have traditionally suffered from the inevitable bias toward higher surface brightness produced by the intensity of the night sky.  Of course, this bias is still a problem, but its effect has been lessened by the advent of more sensitive detectors and the use of larger and better optical systems.  Consequently, in recent years so called low surface brightness (LSB) galaxies have become a topic of great interest.  These previously overlooked members of the galaxy population have since been shown to be larger at the same luminosity as high surface brightness (HSB) galaxies \citep{deb96} while still obeying the same Tully-Fisher relationship, but only if the entire baryonic content of the galaxies is taken into account \citep{mcg00}.  LSB disks also tend to be less dense and more metal poor \citep{deb98} and have higher gas mass fractions than HSB disks \citep{mcg97}.  In addition, attempts at detecting molecular gas in LSB galaxies have been fairly unsuccessful; only 3 out of 34 LSB galaxies observed at the $J$(1-0) and $J$(2-1) CO lines have been found to contain a detectable amount of CO emission \citep{one03}.  Considering these observational trends, it is uncertain whether the characteristics of a sample of galaxies selected according to optical properties would accurately reflect those of the entire galaxy population.\par
Among the galaxy characteristics of particular interest in the optical regime are those that relate to star formation.  Given that the typical LSB galaxy has a metallicity of $\sim \frac{1}{3}$ solar \citep{mcg94}, the lack of detections of CO emission may not be definitive evidence for extremely low amounts of molecular gas in LSB disks, as the CO to H$_{2}$ relation tends to break down for metallicities $\sim \frac{1}{2}$ solar and lower.  Still, the lack of detections, when taken into consideration with LSB disks' low metallicities and H{\sc i} surface densities that are often near or below the critical limit for star formation \citep{deb96,ken89}, imply that these galaxies should not be forming very many stars.  In fact, observations show that the typical LSB disk has a star formation rate that is about ten times lower than the average rate for HSB disks \citep{mih99}.  Despite this, LSB disks tend to be relatively blue and several have been observed to contain a number of prominent H{\sc ii} regions.  The fact that the majority of known LSB disks are bluer might be the result of a selection bias; a fair number of redder LSB galaxies have recently been detected \citep{one00}.  However, the fact that there are so many nearby LSB galaxies that are quite blue along with the presence of H{\sc ii} regions suggests that it is very possible for high mass stars to form in such unfavorable conditions.  This implies that LSB disks may not be characterized simply as less dense and less productive versions of their HSB counterparts and that it is possible that their star formation histories could be much different.\par
A sample of galaxies selected from a single dish H{\sc i} survey will be mainly limited by the total neutral gas content of the galaxies; there is no explicit bias due to surface brightness or H{\sc i} column density.  Therefore, the existence of any fundamental differences between the star formation histories in LSB and HSB galaxies and the relative contribution to the amount of local star formation made by LSB galaxies might be revealed by comparing the star formation properties of a neutral hydrogen selected sample to those of a sample chosen from an optical catalog.  The purpose of this and a subsequent paper is to examine the results of broad band B and R and narrow band H$\alpha$ imaging of a nearby sample of galaxies chosen from the H{\sc i} Parkes All Sky Survey (H{\sc i}PASS) \citep{bar01} in order to derive information about the galaxies' star formation histories and to compare the results with those from an optically selected sample.  This paper discusses the initial results of the data acquisition and reduction, the basic observational properties of each galaxy, and trends among the sample.

\section{Data Acquisition, Reduction, and Calibration}

\subsection{Sample Selection, Observations, and Reduction}

The galaxies that were imaged were chosen from a sample of all H{\sc i}PASS galaxies with declinations $<$ -65$^{\circ}$ and radial velocities $<$ 2500 km s$^{-1}$, 132 galaxies in all.  Due mainly to weather constraints, 69 of these 132 were imaged.  Because of extinction issues, lower priority was given to galaxies that appeared nearly edge-on on their Digitized Sky Survey images available through the NASA/IPAC Extra Galactic Database (NED)\footnote{NED is operated by the Jet Propulsion Laboratory, California Institute of Technology, under contract with the National Aeronautics and Space Administration.}.  Because of this, we poorly sample the more highly inclined portion of the disk population.  The effect of this on our analysis is discussed in Sec. 4.2.  Data was acquired during two separate observing runs.  During both runs, images were obtained at Cerro Tololo Inter-American Observatory (CTIO)\footnote{CTIO is a division of the National Optical Astronomy Observatories, which are operated by the Association of Universities for Research in Astronomy, Inc., under contract with the National Science Foundation.} utilizing the 0.9 m telescope and the Tek 2kx2k CCD camera, one on March 3 - 9, 2000 and the other on Oct. 23-28, Nov. 2, 5, 6, 2000.  Conditions were photometric for all but one of the nights during the first run.  Three of the eight nights of the second run were not photometric; during two of these nights, no images were obtained.  All but three of the H{\sc i}PASS galaxies observed had well known optical counterparts.  Each galaxy was imaged twice through broad band B and R filters and a narrow band H$\alpha$ filter.  In addition to the 3 in., 75 $\mbox{\AA}$ H$\alpha$ filters available at CTIO, 4 in., 30 $\mbox{\AA}$ filters kindly provided by Richard Rand were used during both runs.  Standard IRAF routines in the CCDRED package were used to perform the flat fielding and trimming of the images as well as the overscan corrections.\par

\placetable{tab1}

\subsection{Calibration}

Photometric and spectrophotometric standard stars were imaged during both runs through all filters.  The photometric standard fields were taken from \citet{lan92}; the spectrophotometric standards were obtained from \citet{sto83} and \citet{bal84} (see Table 2 for a list of the fields used as they are identified in these references).  Values for the atmospheric extinction coefficient, $k$, the transformation coefficient, $t$, and the zero point offset, $z$, were obtained by fitting the function $m_{inst}-m_{filter}=k_{filter}X+t_{filter}(B-R)+z_{filter}$ to the observed and published standard star data where $X$ is the airmass of the observation; for this paper, $m_{inst}=-2.5 \mbox{log}(\frac{DN}{\tau})+25$ where $DN$ is the counts from the object and $\tau$ is the exposure time.  For the H$\alpha$ images, the magnitudes of the spectrophotometric standards over the bandpasses of the filters used were calculated using the filter specifications and the magnitude of each star as a function of wavelength, $m_{\lambda}$, quoted in the above references where\par
\begin{equation}
m_{\lambda}=-2.5log(F_{\lambda})-21.1
\end{equation}
\noindent
with $F_{\lambda}$ in units of ergs s$^{-1}$ cm$^{-2}$ $\mbox{\AA}^{-1}$.  These were then compared to the instrumental magnitudes to derive an extinction coefficient and zero point offset for each filter.\par
\placetable{tab2}
Standard star images were only obtained two to three times per night.  Consequently, values for $k$, $t$, and $z$ were obtained for each run instead of each night so that a larger range in airmass values could be used in each fit (see Table 3 for a listing of the values derived for these parameters).  Conditions on the sixth night of the March observing run and the eighth night of the October/November run were not photometric.  The calibration of images taken on these nights is uncertain and the fluxes obtained should be taken as estimates only.  Each galaxy that was observed on one of these nights is flagged with an $\ast$ in Tables 4, 5, and 6.\par
\placetable{tab3}
The galaxy images were corrected for cosmic rays by combining and averaging the pairs of images obtained in each bandpass.  The continuum emission was subtracted from the H$\alpha$ images using the combined R-band images.  To do this, the R-band images were scaled such that on average, the flux from foreground stars match that on the H$\alpha$ images.  No attempt was made to correct for the presence of emission lines in the R bandpass for the galaxies.  Therefore, this method may slightly overestimate the level of galaxy continuum emission on each H$\alpha$ image.  However, the emission line equivalent widths for typical star forming disk galaxies lie between about 10 and 100 $\mbox{\AA}$ \citep[e.$\:$g.][]{ken83}, implying that for the R-band filter which has a FWHM of $\sim$1000 $\mbox{\AA}$, the presence of emission lines will cause the continuum level to be off by only about 1 to 10\% which is less than or approximately equal to the typical uncertainty for the zero points for the H$\alpha$ filters ($\sim$0.1 mag.).\par
Following sky subtraction, for each galaxy the B-R color was solved for using an estimate of the B and R instrumental magnitudes according to the following, 
\begin{equation}
B-R= \frac{(m_{B}-m_{R})_{inst}-(k_{B}X_{1}-k_{R}X_{2})-(z_{B}-z_{R})}{1+t_{B}-t_{R}}
\end{equation}
\noindent
Each combined B and R and continuum subtracted H$\alpha$ image was then then divided by its exposure time and multiplied by 10$^{0.4[k_{filter}X+t_{filter}(B-R)+z_{filter}-25]}$ where $t_{H\alpha}$=0.\par
Due to the larger number of H$\alpha$ filters used, fewer spectrophotometric standard star observations were made through each filter on the sixth night of the October/November run than on previous nights, and no reasonable results for the calibration coefficients could be obtained.  In addition, only the 75$\mbox{\AA}$ CTIO H$\alpha$ filters were available during the previous nights whereas on the sixth night, the 30 $\mbox{\AA}$ Rand filters were used.  Therefore, the calibration coefficients for the previous nights could not be used, and the calibration of the H$\alpha$ data for this night was performed using the R-band data in the following manner.  It was assumed that the average monochromatic flux across the R filter, $F_{R}$, is proportional to the average monochromatic flux of the continuum across the H$\alpha$ filter, $F_{c}$, according to $F_{R}=C \cdot F_{c}$, which gives the following,\par
\begin{equation}
m_{c}=m_{R}+2.5log(C)+0.5
\end{equation}
\noindent
where $m_{R}$ is the magnitude in the R-band, $m_{c}=-2.5log(F_{c})-21.1$, and the average monochromatic flux and magnitude of Vega in the R-band are taken to be 2.15$\times$10$^{-9}$ ergs s$^{-1}$ cm$^{-2}$ $\mbox{\AA}^{-1}$ and 0.07 respectively.  The raw counts from the R-band image, $DN_{R}$, are related to the counts from the continuum across the H$\alpha$ filter, $DN_{c}$, by a scale factor, $a$, that is determined during the continuum subtraction process, i.$\:$e. $DN_{R}=a \cdot DN_{c}$.  The difference between the calibrated and instrumental magnitudes, $\Delta m$, for the H$\alpha$ filter can then be related to that for the R filter by the following,\par
\begin{equation}
\Delta m_{H\alpha} = \Delta m_{R} + 2.5log(C \cdot a \cdot \frac{\tau_{R}}{\tau_{H\alpha}}) + 0.5
\end{equation}
\noindent
where $\tau$ is the exposure time.  A mean value of $C$=0.73$\pm$0.15(1$\sigma$) was derived from the data from both runs that were calibrated with standard stars (see Fig. 1); a value of $C$=0.73 was used in the calibration of the remaining data.  Given that all the data from the eighth night of the October/November run was calibrated using the standard star data from the sixth night of that run, the H$\alpha$ fluxes for galaxies imaged on that night were calibrated in this manner as well.  Each galaxy whose H$\alpha$ flux was derived in this way is denoted by a $\star$ in table 6.\par

\begin{figure}
\plotone{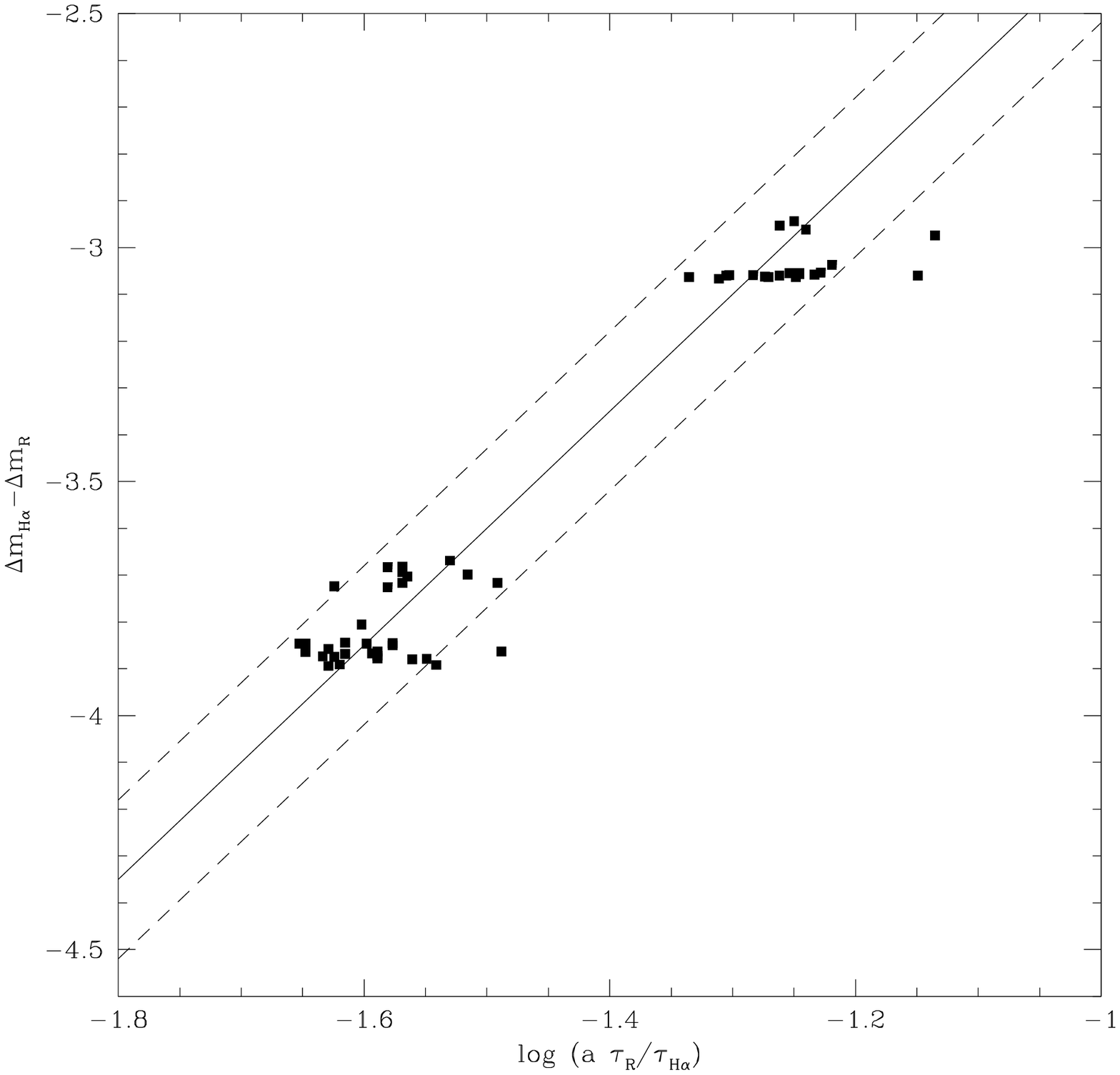}
\caption{The difference in the H$\alpha$ and R-band calibrations as a function of continuum subtraction scaling factor, $a$, and exposure times for all data calibrated with standard stars.  The solid and dashed lines are the mean value for $C$ $\pm 2 \sigma$ (see Sec. 2.2) used to compute the H$\alpha$ calibration from the R-band data for galaxies imaged during the final two nights of the second observing run (see Sec. 2.2 for a more detailed discussion).  The two discrete larger groups of points correspond to the two different types of H$\alpha$ filters used, the 30 $\mbox{\AA}$ Rand filters (lower group) and 75 $\mbox{\AA}$ CTIO filters (upper group).  Within each larger group, there are two smaller discrete groups; for the Rand filters, the upper group corresponds to the 657 nm filter and the lower group corresponds to the 660 and 661 nm filters; for the CTIO filters, the upper group corresponds to the calibration for the first observing run and the lower group corresponds to the calibration for the second run.}
\label{fig1}
\end{figure}

\placefigure{fig1}

\section{Photometric Measurements}

\subsection{Isophotal and Surface Photometry}

After the combined B and R and H$\alpha$ continuum subtracted images were constructed and calibrated, isophotal photometry was performed on the galaxies in the following manner.  On the calibrated B-band image, the 25 mag arcsec.$^{-2}$ isophote was identified by smoothing the image with a 7 pixel wide boxcar and then locating all pixels within 3$\sigma_{cal}$ of 25 mag arcsec.$^{-2}$, where $\sigma_{cal}$ is the uncertainty in the flux due to the error in the calibration.  The width of the boxcar was chosen to be big enough to remove noise features from each image without eliminating the basic shape of the galaxy's surface brightness profile (7 pixels corresponds to about 0.5 kpc for $V_{R}$=2500 km s$^{-1}$).  These pixels were fit with an ellipse to determine the center, ellipticity, and position angle of the 25 mag arcsec.$^{-2}$ isophote; a polygon aperture was then drawn around these pixels to isolate the galaxy on the image.  A pixel mask was created by replacing all pixels with fluxes greater than or equal to 25 mag arcsec.$^{-2}$ with values of unity and the rest with values of zero.  For each galaxy, all three images were multiplied by this mask and the fluxes within the polygon aperture for the resulting images were measured to obtain isophotal magnitudes in all three bands.  The mask was then used to estimate the isophotal radius for the object; this radius was taken to be the radius of a circle with the same area as the space occupied by all of the non-zero pixels on the mask image that were within the polygon aperture.  These radii are reported in Table 4 along with estimates for the position angles and inclinations for the galaxies based on the shapes of the 25 mag. arcsec.$^{-2}$ isophotes.  The isophotal magnitudes are contained in Table 4 as well.  For each measured flux, the uncertainty was computed and was found to be dominated by the uncertainty in the calibration coefficients; the mean error for the fluxes is approximately the same for both bands and is equal to $\sim$0.045 mag.\par
For one galaxy, ESO035-G009, the observed surface brightness in the center of the object on the B-band image was approximately 25 mag arcsec.$^{-2}$.  Therefore, a 25 mag arcsec.$^{-2}$ isophote could not be identified.  The B-band image was sufficiently deep (the exposure time used was twice the typical time used for the rest of the galaxies) that a 26 mag arcsec.$^{-2}$ isophote could be identified.  Therefore the isophotal quantities reported for this galaxy are for the 26 mag arcsec.$^{-2}$ isophote; the galaxy is flagged with a $\dag$ in Table 4.\par
\begin{figure}
\plotone{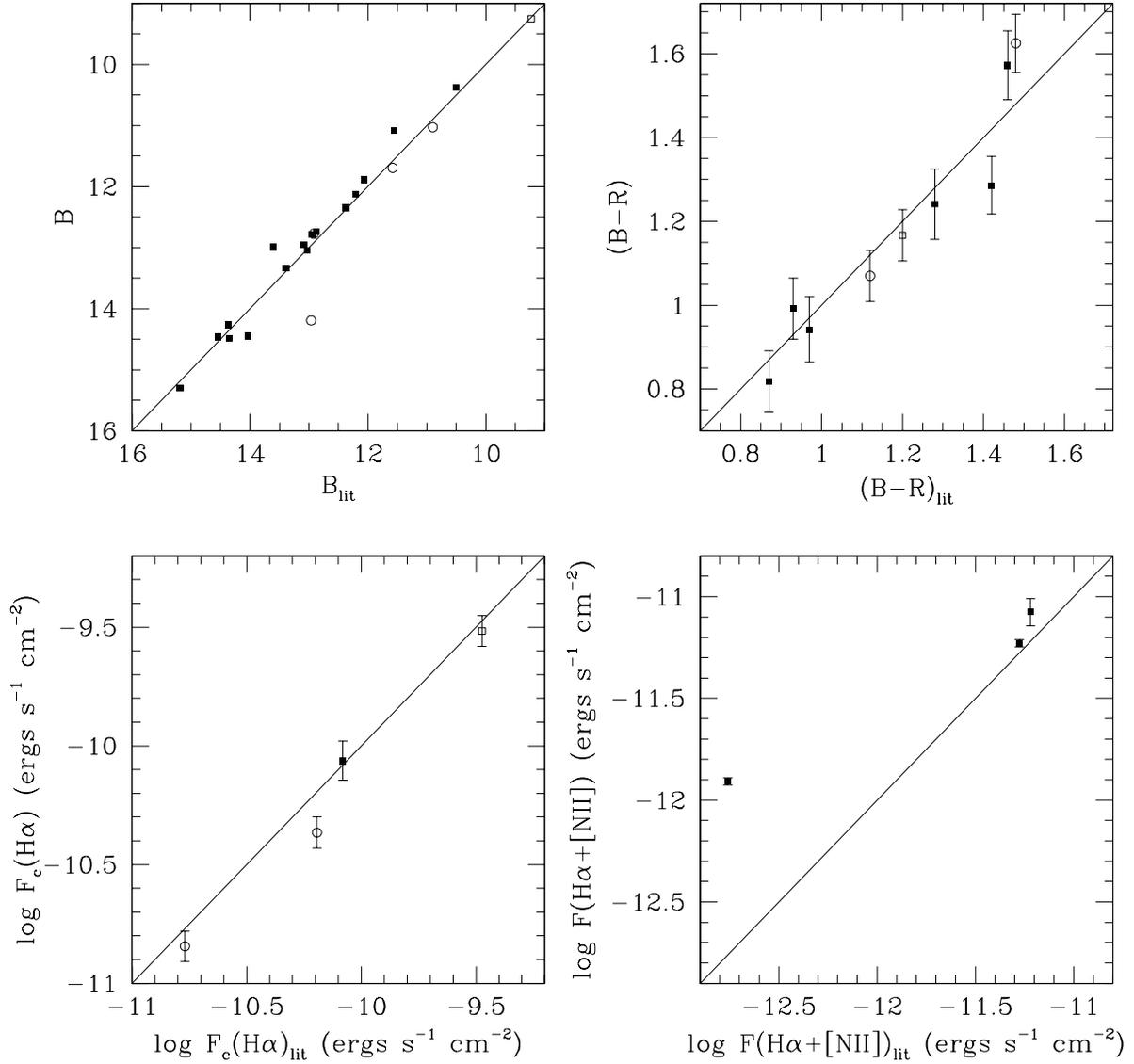}
\caption{Comparison of the photometry performed on the H{\sc i}PASS sample and that found in the literature (see Sec. 3.1 and 3.2 for references) for B-band magnitudes (upper left), apparent (B-R) color indexes (upper right), H$\alpha$ fluxes corrected for extinction and $[$NII$]$ contamination (lower left), and H$\alpha+[$NII$]$ fluxes (lower right).  The open circles are data derived from questionable calibration (see Sec. 2.2); the open box is M83; the error bars represent the 1$\sigma$ errors in the photometry.  In the lower right panel, the point that deviates from the published value by 0.85 dex is discussed in Sec. 3.2.}
\label{fig2}
\end{figure}

\placefigure{fig2}
\placetable{tab4}
Following this, B and R surface brightness profiles were measured.  Due to the unusual shape of the inner isophotes of many Sm and Im galaxies (see Fig. 3 for an example), elliptical isophotal fitting proved difficult for many of these objects.  Therefore, the profiles were obtained by measuring the median flux within successive elliptical annuli, each with the center, ellipticity, and position angle of the B-band 25 mag arcsec.$^{-2}$ isophote.  The radius of each annulus was taken to be $\sqrt{ab}$ where $a$ and $b$ are the semi-major and semi-minor axes of the ellipse half way in between the boundaries of the annulus.  The radius of the inner most annulus was chosen to be half of the typical FWHM of the seeing (1.5 arcsec.); the outer radius was set to approximately twice the isophotal radius.  The spacing between successive pairs of annuli was set to be 1.1 times larger than the spacing for the previous annuli to maintain a nearly constant signal to noise ratio.\par
The resulting B and R profiles were simultaneously fit with an exponential profile for all radii where the B-band surface brightness was between 0.3 and 2.472 mag above the 1$\sigma$ limiting isophote (i.$\:$e. two disk scale lengths for a pure exponential disk).  A standard weighted, nonlinear least-squares fitting routine was used to perform the fits and to obtain uncertainties in the derived quantities, namely the central disk surface brightness values, $\mu_{o,B}$ and $\mu_{o,R}$, and the disk scale length, $h$.  The median error is approximately 0.07 mag. for the central surface brightness values and 2.5\% for the scale lengths.  Examples of measured profiles and disk fits are displayed in Fig. 4.  From these examples, it can be seen that the above procedure underestimates the true surface brightness of bars or bulges that do not have the same ellipticity and/or orientation as the the disks that contain them.  However, the measured surface brightness values for the disk components are more relavent for deriving properties for these objects (e.$\:$g. disk scale lengths, central surface brightness values, extrapolated fluxes) and many of the objects have no discernible central component.\par
The disk fits were used to obtain total extrapolated fluxes in both bands.  Uncertainties were computed for these extrapolated values using the errors for the measured fluxes and those for the disk profile parameters; the typical error is $\sim$0.1 mag.  The B-band profile was then interpolated to find the radius that contained half of this total extrapolated flux; this was taken to be the effective radius, $r_{e}$.  The B-R color within $r_{e}$ and the B and R surface brightness values at $r_{e}$ were then measured.  The error in these quantities was estimated for each galaxy in the following way.  The half-light radius was determined for 100 different values of $B_{T}$ ranging from $B_{T}-2 \cdot \sigma_{B_{T}}$ to $B_{T}+2 \cdot \sigma_{B_{T}}$, where $\sigma_{B_{T}}$ is the computed error in the extrapolated B-band magnitude.  For each new value of $r_{e}$, the color and surface brightness values were measured and a weight was computed using the corresponding total B-band flux and a Gaussian function with a mean equal to the measured value for $B_{T}$ and a standard deviation equal to $\sigma_{B_{T}}$.  For each of these quantities, a weighted mean was then computed, and the error for each quantity was then taken to be the weighted standard deviation about the mean.  The typical values for the error in $r_{e}$, $\mu_{e}$, and $(B-R)_{e}$ are approximately 10, 20, and 12\% respectively.  All surface photometry parameters are listed in Table 5.; all fluxes and surface brightness values in Table 5 are corrected for Galactic extinction using E(B-V) values obtained from NED and E(B-V) to $A_{filter}$ conversion factors from \citet{schl98}.\par
To check our broad-band photometry, we use the catalog of galaxy photoelectric photometry compiled by \citet{pru98}.  In that catalog, 20 of our galaxies have B-band magnitudes reported; 9 have R-band fluxes.  The catalog also reports the size of the circular aperture used for each measurement.  To properly compare our measurements, we measured the fluxes from our calibrated B and R images using the same size circular apertures; the center of each aperture was chosen to be the center of the best fitting ellipse for each galaxy's 25 mag. arcsec.$^{-2}$ isophote.  The results are plotted in Fig. 2.  In general, our B-band fluxes are in agreement (the median deviation is $\sim$0.13 mag).  The most notable discrepancy involves IC2554 for which our measured flux is about 1.2 mag. fainter than the published value.  The discrepancy is most likely due to the fact that the images for this galaxy were taken during intermittent cloudy conditions.  However, it should be noted that fluxes for two other galaxies that were imaged in similar conditions agree quite well with the values from the literature.  In general, the B-R colors we have measured for our galaxies agree with the published values within 2$\sigma$ (median deviation of $\sim$0.05 mag.).\par

\begin{figure}
\plotone{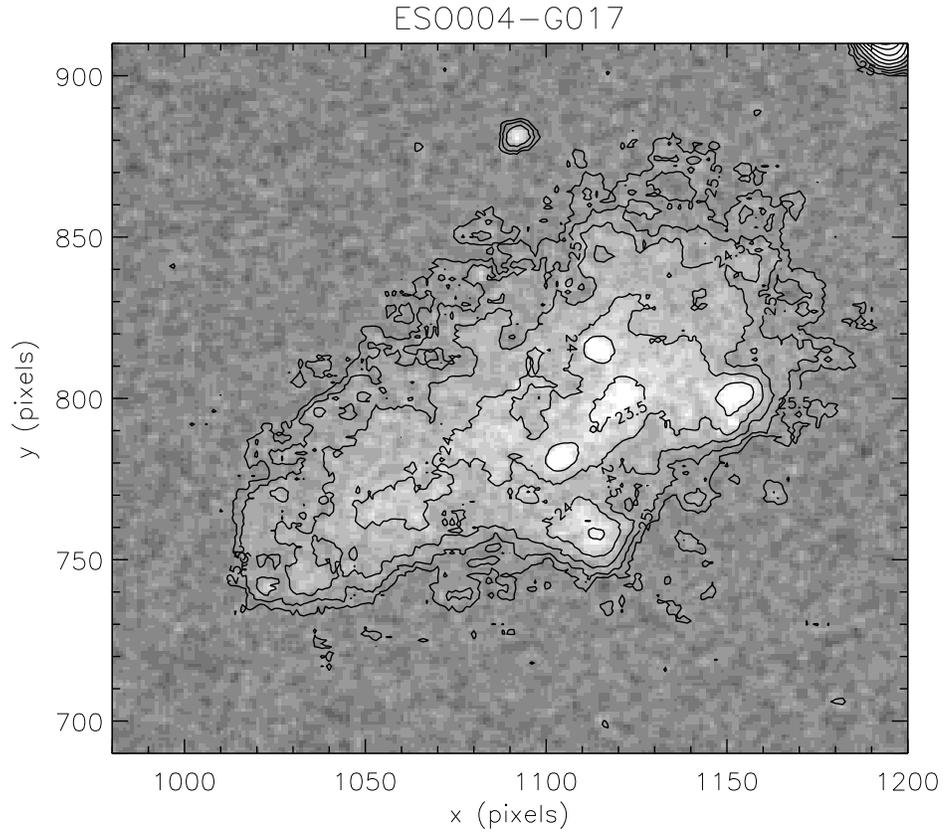}
\caption{B-band image of ESO004-G017, an Sm galaxy, with contours drawn for different isophotes; the surface brightness of each isophote is given in mag arcsecond$^{-2}$.  Note that there are three discrete 23.5 mag arcsecond$^{-2}$ isophotes caused by irregularly distributed H{\sc ii} regions; the 24 mag arcsecond$^{-2}$ isophote is broken up into distinct components as well.}
\label{fig3}
\end{figure}

\begin{figure}
\plotone{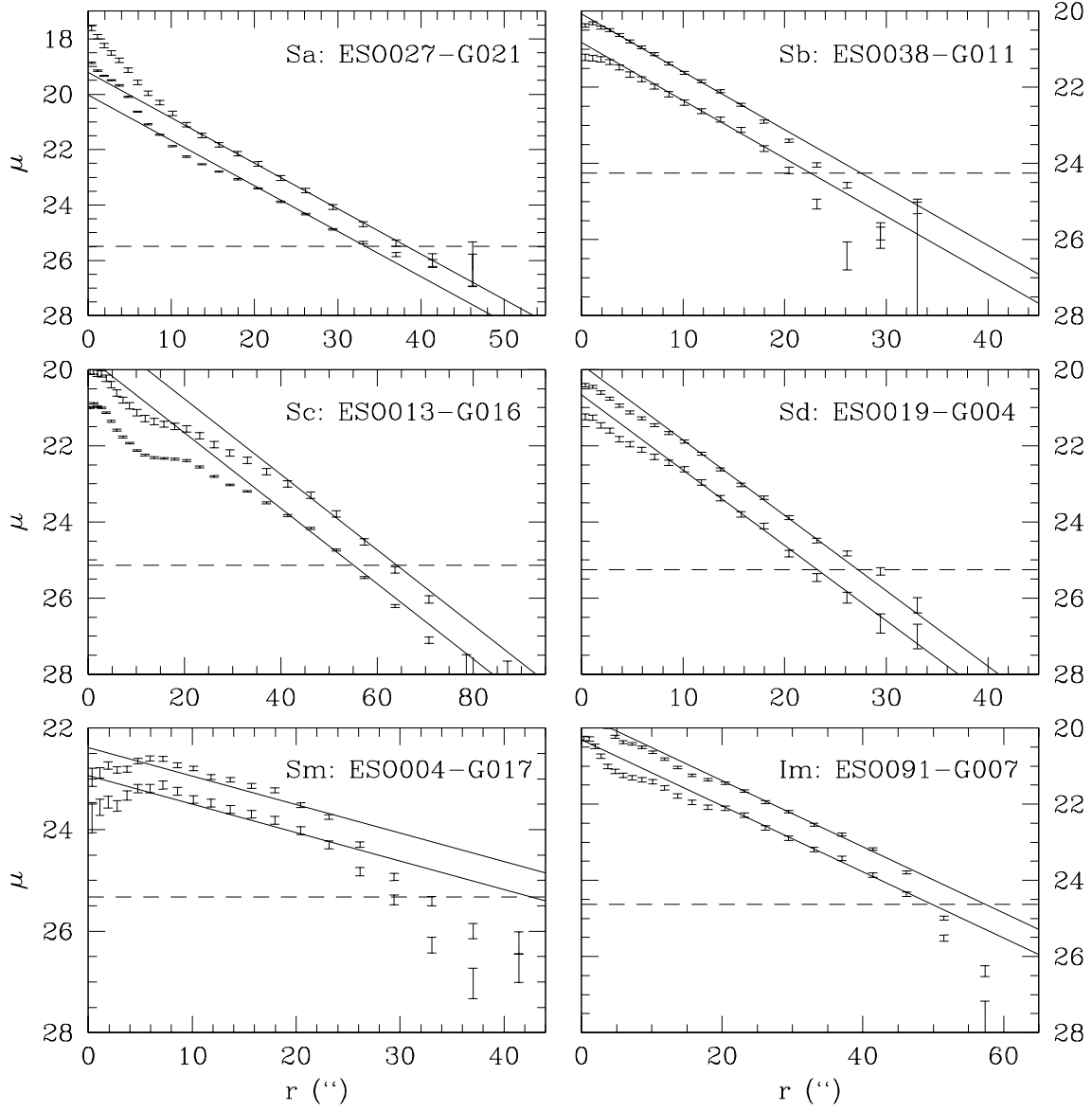}
\caption{Example surface brightness profiles for 6 disk morphologies along with the exponential fits; the lower points are the B-band profiles; the upper points are the R-band profiles; all values of $\mu$ are corrected for Galactic extinction; all radii are in units of arcseconds.  The horizontal dashed lines represent the B-band 1$\sigma$ limiting isophote (corrected for Galactic extinction).}
\label{fig4}
\end{figure}

\placefigure{fig3}
\placefigure{fig4}
\placetable{tab5}

\subsection{H$\alpha$ Fluxes and Equivalent Widths}

The H$\alpha$ flux for each object was computed using the calibrated H$\alpha$ isophotal magnitude and the appropriate filter transmission curve according to the following,\par
\begin{equation}
F_{H\alpha}=\int F_{l} d \lambda = \frac{\int T_{H \alpha} d \lambda}{T_{H\alpha}( \lambda_{l} )}10^{-0.4(m_{H\alpha}+21.1)}
\end{equation}
\noindent
where $F_{l}$ is in units of ergs s$^{-1}$ cm$^{-2}$ $\mbox{\AA}^{-1}$, $T_{H\alpha}$ is the transmission of the H$\alpha$ filter, $m_{H\alpha}$ is the isophotal magnitude measured from the calibrated continuum subtracted H$\alpha$ image, and $\lambda_{l}$ is the location of the H$\alpha$ emission line (=6563$\mbox{\AA} \cdot$(1+z)) determined using the redshift measured from the H{\sc i}PASS 21 cm data.\par
To remove the the effects of contamination from $[$NII$]$ emission and internal extinction, data for the Nearby Field Galaxy Survey (NFGS) of \citet{jan00} was used to derive relations between R-band luminosity and $[$NII$]$/H$\alpha$ and the internal extinction at H$\alpha$, $A(H\alpha)_{int}$ (for more details on the NFGS, see Sec. 4.1).  Using a linear least squares fit, the measured ratios of $[$NII$]$ to H$\alpha$ reported by \citet{jan00} yielded the following relation,\par
\begin{equation}
\mbox{log }\frac{[NII]}{H\alpha}=[-0.13\pm0.035]M_{R}+[-3.2\pm0.90]
\end{equation}
\noindent
where $M_{R}$ is the absolute magnitude in the R-band.  The total amount of extinction for each NFGS galaxy with a measured H$\beta$ flux was computed using the values for the ratio of H$\alpha$ to H$\beta$ measured by \citet{jan00}, an assumed intrinsic ratio of $\frac{H\alpha}{H\beta}$=2.85 (for case B recombination and T=10$^{4}$ K \citep{ost89}), the extinction curve of \citet{odo94}, and $R_{V}$=3.1.  A value for E(B-V) obtained from NED for each of these galaxies was then used to compute the Galactic extinction at H$\alpha$ according to the same extinction law; this was subtracted from the total extinction computed using the Balmer decrement to generate a value for $A(H\alpha)_{int}$.  These values were used with a linear least squares fitting routine to obtain the following,\par
\begin{equation}
\mbox{log }A(H\alpha)_{int}=[-0.12\pm0.048]M_{R}+[-2.5\pm0.96]
\end{equation}
\noindent
Given that the NFGS spans approximately the same range in luminosity as our H{\sc i}PASS sample (see Fig. 9), these relations were deemed adequate for correcting the H$\alpha$ fluxes for our H{\sc i}PASS galaxies.  The H$\alpha$ fluxes for the H{\sc i}PASS galaxies were also corrected for Galactic extinction using E(B-V) values from NED, the extinction curve of \citet{odo94}, and $R_{V}$=3.1.\par
The median error in these fluxes due to the photometry (i.$\:$e. not including the uncertainty in the above mentioned corrections) is $\sim$13\%.  The H$\alpha$ fluxes, corrected using the equations above, are reported in Table 6.  For the determination of the H$\alpha$ equivalent widths, the continuum flux was assumed to be constant across the line.  The mean monochromatic continuum flux was determined by measuring the fluxes from the calibrated H$\alpha$ and H$\alpha$ continuum subtracted images and taking the difference between the two.  The H$\alpha$ emission line equivalent width was then taken to be the integrated line flux corrected for $[$NII$]$ contamination divided by this mean monochromatic continuum flux.  These values are also listed in Table 6.\par 
Three of our galaxies (NGC1313, NGC2442, and NGC5068) have H$\alpha$ fluxes reported in \citet{ryd94}.  The fluxes for these galaxies were corrected for contamination from $[$NII$]$ emission by \citet{ryd94} using the typical ratios for $[$NII$]$/H$\alpha$ for spirals of 0.33 and for Sm and later galaxies of 0.053 reported in \citet{ken83}.  \citet{ryd94} also corrected the fluxes for 1.1 mag of internal extinction as recommended by \citet{ken83} and for Galactic extinction assuming $A_{H\alpha}=0.64A_{B}$.  An H$\alpha$ flux for M83 that was obtained from narrow-band photometry and corrected for $[$NII$]$ contamination and Galactic extinction was also taken from \citet{bel01}; an extra correction for 1.1 mag of internal extinction was applied to this published flux for M83.  For comparison, we have applied these same corrections to our H$\alpha$+$[$NII$]$ fluxes for NGC1313, NGC2442, NGC5068, and M83; the results are displayed in the lower left panel of Fig. 2.  Our corrected fluxes agree with the published values within 2$\sigma$.\par
Uncorrected narrow-band fluxes were also obtained for three of our objects (NGC1511, NGC6300, and NGC7098) from \citet{cro96}.  We compare our uncorrected H$\alpha+[$NII$]$ fluxes with these published values in the lower right panel of Fig. 2.  Again, the fluxes agree within 2$\sigma$ with one obvious exception, NGC7098.  Examination of the H$\alpha$ continuum subtracted image for this galaxy displayed in \citet{cro96} along with the reported ratio of the H{\sc ii} region to total H$\alpha$ flux revealed that there may be a significant discrepancy between our continuum subtraction and that performed by \citet{cro96}.  The value reported by the authors for this ratio is 76\% whereas an estimate from our continuum subtracted image places the fraction at less than 20\%.   Therefore, the discrepancy in flux most likely arises from a difference in continuum subtraction, not calibration.  It should be noted, however, that fluxes from the other two galaxies in our sample that were observed by \citet{cro96} are in good agreement and that the continuum subtraction procedure used for those objects was identical to that used for NGC7098.\par

\placetable{tab6}

\section{Results and Discussion}

\subsection{Comparison Samples}
As an optically selected comparison sample, we have chosen the Nearby Field Galaxy Survey (NFGS) of \citet{jan00}.  This is a spectrophotometric sample of 198 galaxies from the CfA redshift survey \citep{huc83} that spans a wide range in both luminosity and morphology.  The survey area is that of the CfA survey, two circular regions centered on the north and south Galactic poles with radii of 50$^{\circ}$ and 60$^{\circ}$ respectively, except that galaxies within 6$^{\circ}$ of the center of the Virgo cluster out to V$_{R} \leq$2000 km s$^{-1}$ were not included.  The sample was chosen from a magnitude limited subsample of 1006 galaxies that excluded galaxies with V$_{LG}>$10$^{-0.19-0.2M_{z}}$ where V$_{LG}$ is the radial velocity relative to the Local Group and $M_{z}$ is the absolute photographic B magnitude from the Zwicky catalog.  Within this subsample, the galaxies were binned in $M_{z}$.  Within each bin, the galaxies were sorted by morphology and ever $N$th galaxy was selected from each bin where $N$ is the ratio of the number of galaxies in the bin to the desired number.  For each bin, the desired number was chosen based on the local galaxy luminosity function.  The maximum radial velocity for the resulting sample of 198 galaxies is $\sim$11,000 km s$^{-1}$ (only four galaxies are beyond this limit).  Optical spectra and UBR images were obtained by \citet{jan00} for all 198 galaxies.  \citet{jan00} also measured equivalent widths for all detected emission lines and performed surface UBR photometry.\par
\begin{figure}
\plotone{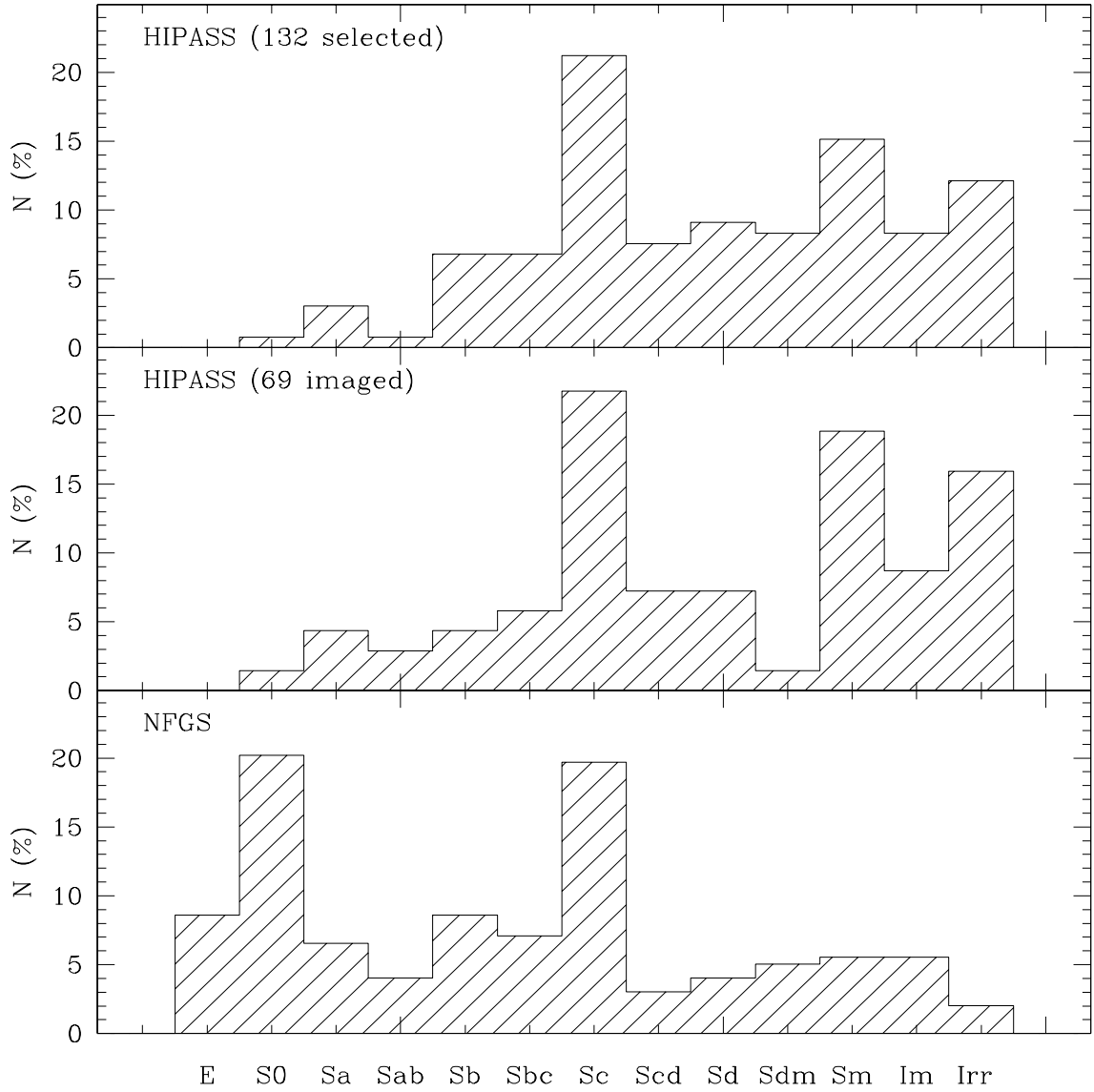}
\caption{The distribution of morphological types among the 132 H{\sc i}PASS galaxies initially selected (top panel), H{\sc i}PASS galaxies included in this study (middle panel), and the NFGS galaxies (bottom panel).}
\label{fig5}
\end{figure}

\placefigure{fig5}
\citet{jan00} used a procedure nearly identical to the one described in Sec. 3.1 to obtain extrapolated total fluxes and effective radii from all measure surface brightness profiles.  To obtain disk parameters and R-band extrapolated fluxes, we applied our procedure to the profiles made publicly available\footnote{currently available at http://cfa-www.harvard.edu/$\sim$jansen/nfgs/nfgs.html} by \citet{jan00}; we calculate values for $B_{T}$, $r_{e}$, and (B-R)$_{e}$ that are in excellent agreement ($\sim$1\% deviation) with those reported by \citet{jan00}.  A follow-up paper by \citet{kew02} on the comparison between H$\alpha$ and IR derived SFRs contains calibrated H$\alpha$ fluxes for 93 NFGS galaxies that were matched with IRAS sources.  These fluxes were used to calculate SFRs which we compare with those for the H{\sc i}PASS galaxies.  Fluxes at 21 cm were obtained for 87 NFGS galaxies from the RC3 catalog for evaluation of their gas content.\par
\begin{figure}
\plotone{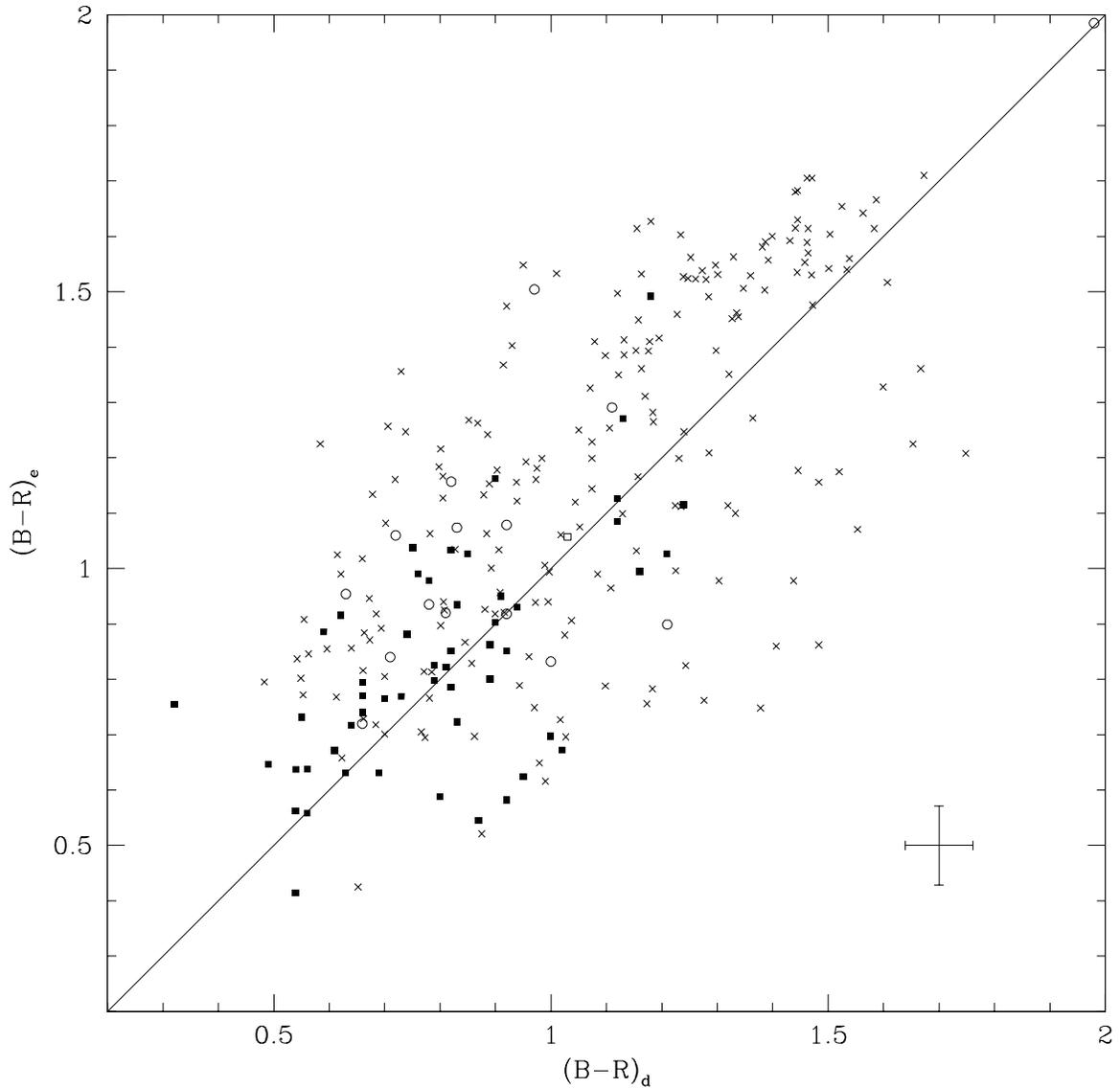}
\caption{B-R color derived from the exponential disk parameters versus the B-R color within the effective radius for the H{\sc i}PASS and NFGS galaxies.  NFGS galaxies are represented by $\times$'s; closed boxes represent H{\sc i}PASS galaxies; open circles represent H{\sc i}PASS galaxies for which the calibration is questionable; the open box is M83.  All colors are corrected for Galactic reddening.  The displayed error bars are the median 1$\sigma$ errors for the H{\sc i}PASS galaxies.  The line plotted is what would be expected if the disk and effective radius colors were the same; the points do not follow this trend as the majority of the points for both samples have disk colors that are bluer than their effective radius colors.}
\label{fig6}
\end{figure}
\begin{figure}
\plotone{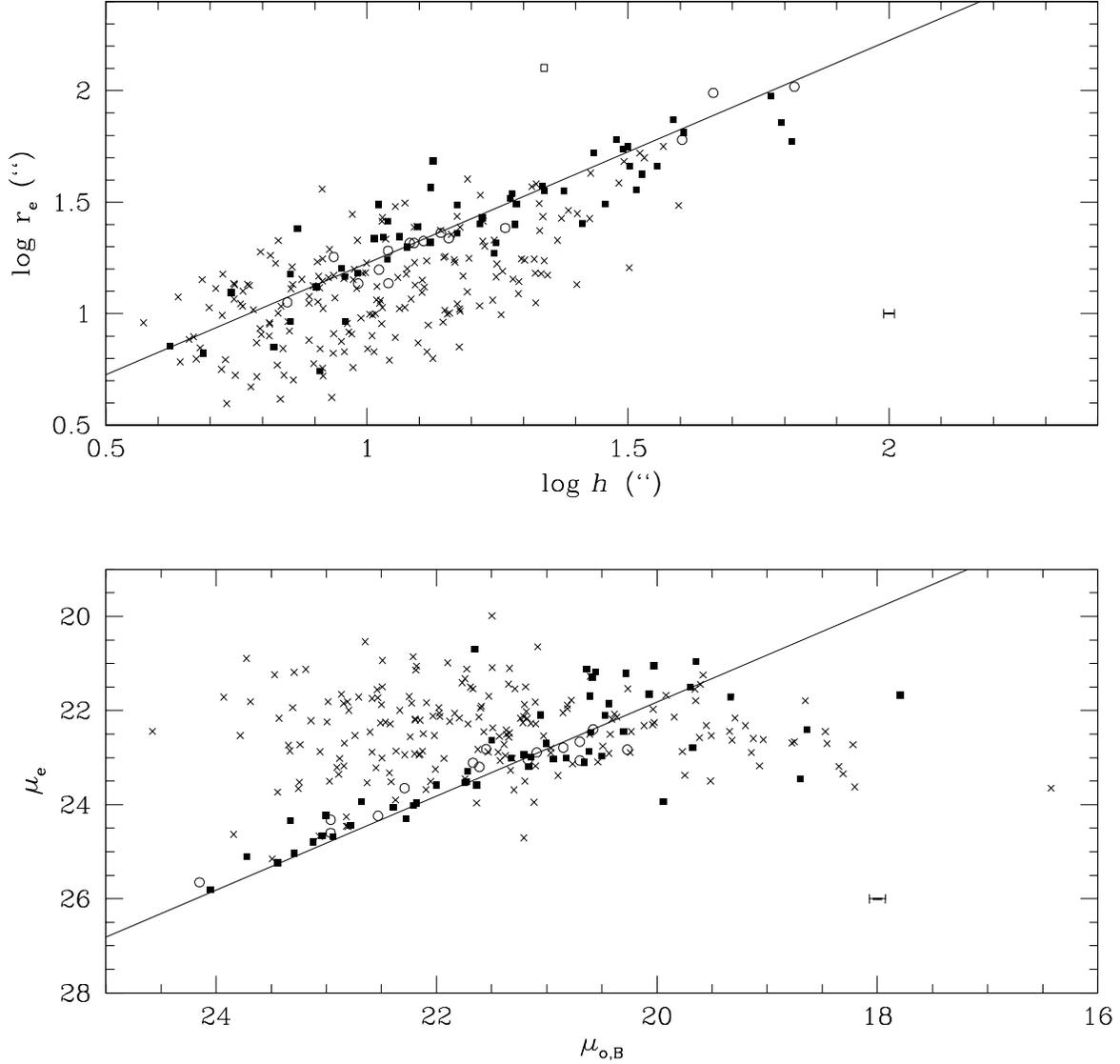}
\caption{Disk scale length versus effective radius (upper panel) and central disk surface brightness versus surface brightness at the effective radius (lower panel).  NFGS galaxies are represented by $\times$'s; closed boxes represent H{\sc i}PASS galaxies; open circles represent H{\sc i}PASS galaxies for which the calibration is questionable; the open box is M83.  The displayed error bars are the median 1$\sigma$ errors for the H{\sc i}PASS galaxies.  The lines in both panels represent the relations expected for pure exponential disks; the H{\sc i}PASS galaxies follow these trends well whereas the NFGS galaxies do not.}
\label{fig7}
\end{figure}

\placefigure{fig6}
\placefigure{fig7}
We also compare the gas content as a function of optical properties for the H{\sc i}PASS galaxies to the results of a study performed by \citet{scho01}.  The purpose of their study was to explore whether or not the trends observed among ``normal'' spirals extended to LSB dwarfs.  The data for ``normal'' spirals was take from samples of \citet{cou96} and \citet{dej96}.  The LSB dwarfs were chosen from the Second Palomar Sky Survey plates and observed at 21 cm with the Arecibo 305 m telescope out to a radial velocity of 8120 km s$^{-1}$ and in the optical with the Hiltner 2.4 m telescope located at Michigan-Dartmouth-MIT (MDM) Observatory.\par

\subsection{Overall Properties}
Fig. 5 shows that as suspected, the H{\sc i}PASS sample contains a much higher fraction of late-type disks than the optically selected NFGS.  Fig. 6 demonstrates that for both samples, disk colors calculated with the extrapolated central surface brightness values, $\mu_{o}$, derived from the exponential fits are bluer that the (B-R)$_{e}$ colors for the majority of the galaxies in both samples ($\sim$70\% for both).  However, Fig. 7 shows that the NFGS galaxies deviate from the relations between $r_{e}$ and disk scale length, $h$, and $\mu_{e}$ and $\mu_{o}$ expected for a pure exponential disk to a significantly larger degree than the H{\sc i}PASS galaxies.\par
The optical and H{\sc i} properties of the H{\sc i}PASS sample are displayed in Fig. 8, 9, and 10 along with similar histograms for the NFGS.  All distance dependent quantities were calculated using $h$=0.70 and radial velocities that were corrected to the rest frame of the Local Group according to V$_{LG}$=V$_{R}$+$\Delta$V$_{Virgo}$+300$cos(b)sin(l)$ \citep{jan00}, where $\Delta$V$_{Virgo}$ is the correction applied to the observed radial velocity, V$_{R}$, for Virgo-centric infall.  The histograms indicate that the H{\sc i}PASS sample contains a higher fraction of bluer, more LSB galaxies than the NFGS sample.  A Kolmogorov-Smirnov (K-S) test was performed on each pair of histograms, the results of which are listed in Table 7.  For the majority of the properties explored, the probability that the histograms for the two samples come from the same parent distribution are low.  This is not surprising since the NFGS was chosen to span a wide range in galaxy types.  However, the probabilities for surface brightness at the effective radius and color indexes are orders of magnitude lower than those for the other histogram pairs.  The fact that edge-on galaxies were excluded from the H{\sc i}PASS sample may contribute to the large discrepancy in the color histograms (the probability from the K-S test is $\sim 10^{-10}$).  However, Fig. 11 demonstrates that for all values of inclination (or ellipticity), the H{\sc i}PASS galaxies are preferentially bluer and that the more highly inclined NFGS galaxies are not significantly redder than the rest of the sample.  Therefore, the effect of excluding edge-on galaxies on the difference in color distributions is likely minimal.\par
To test how the higher fraction of elliptical and lenticular galaxies contained within the NFGS effects these results, the K-S tests were run again using only Sa or later type galaxies.  These values are also listed in Table 7.  It can be seen that extremely low probability for the distributions of color is most likely caused by early-type galaxies, as the probability for spirals is about 4.5 orders of magnitude higher.  However, the probability is still $\sim$10$^{-6}$.  The exclusion of early-type galaxies increases the probability that the two surface brightness distributions come from the same parent distribution by a factor of about 10.  But, again, the probability is still quite low ($\sim$10$^{-4}$).  This implies that a galaxy sample similarly selected to cover a larger variety of morphologies from a parent sample taken from an H{\sc i} survey may contain a higher fraction of bluer, lower surface brightness galaxies than if the parent sample was taken from an optical catalog.\par
\begin{figure}
\plotone{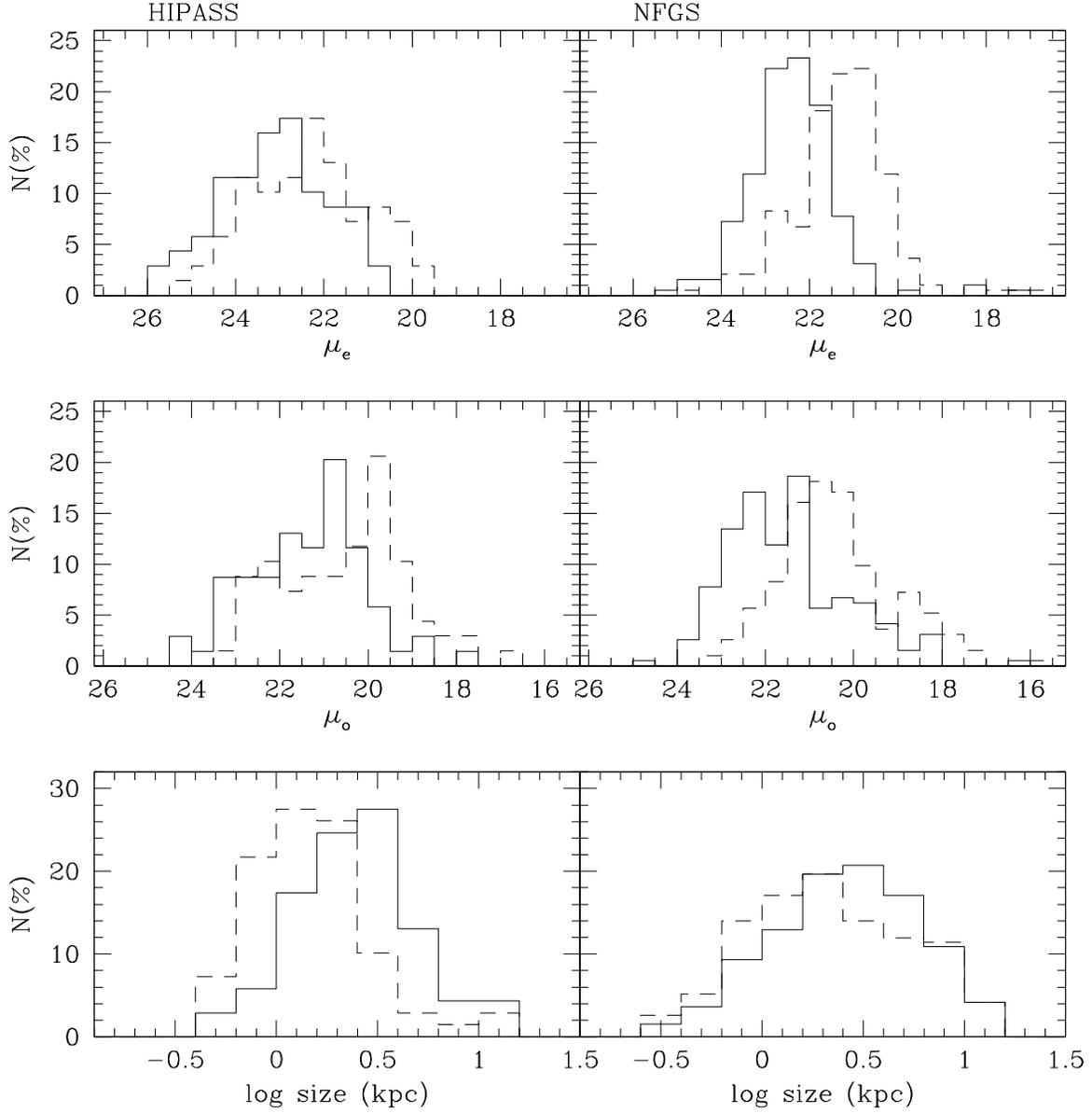}
\caption{The properties of the H{\sc i}PASS sample pertaining to the distribution of stellar light (surface brightness at the half-light radius, $\mu_{e}$, central disk surface brightness, $\mu_{o}$, half-light radius, $r_{e}$, and disk scale length, $h$) along with similar histograms for the NFGS galaxies; the solid and dashed lines represent B-band (effective radius) and R-band (disk scale length) data respectively; distance dependent quantities are computed with $h$=0.70.}
\label{fig8}
\end{figure}
\begin{figure}
\plotone{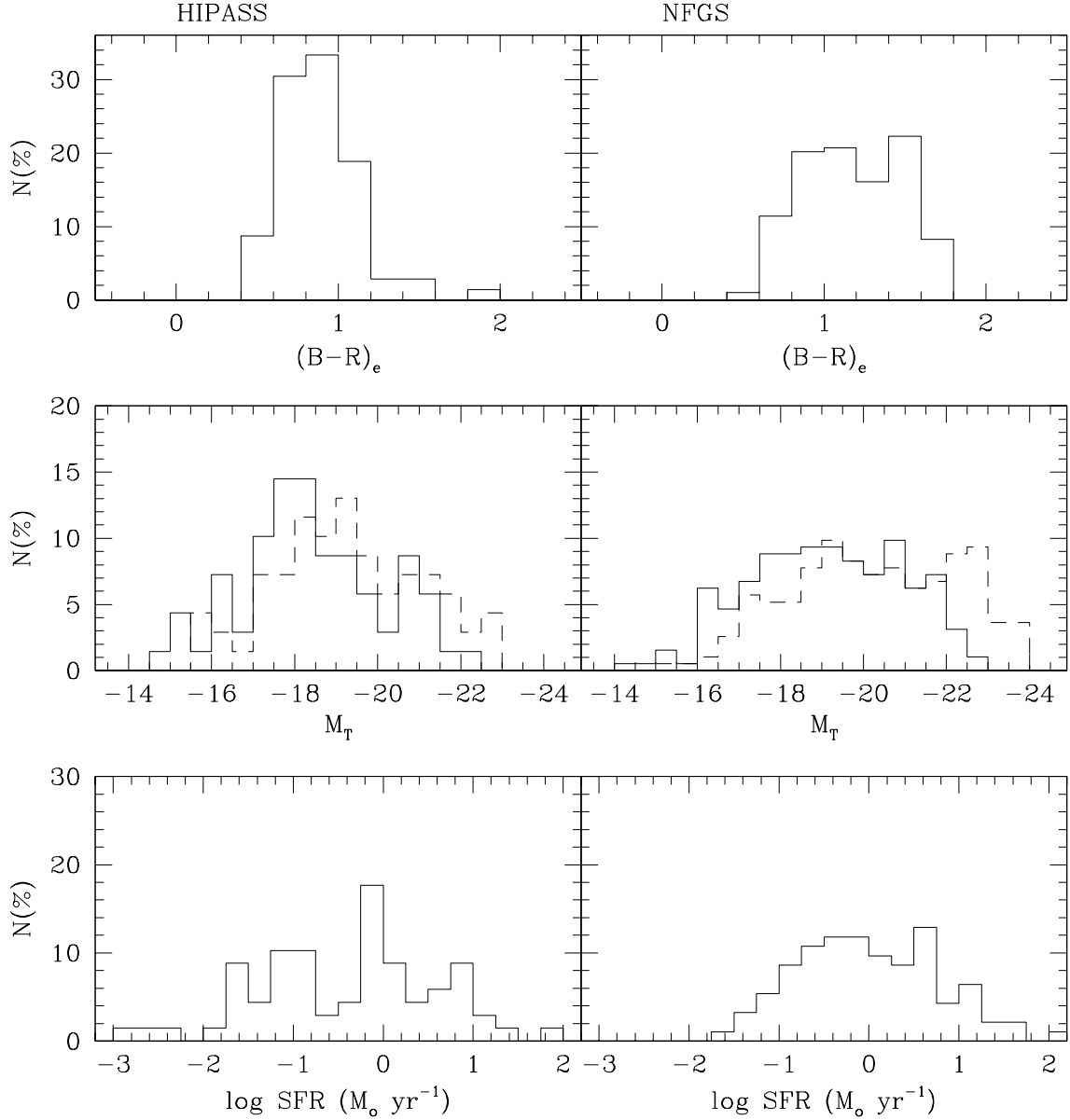}
\caption{The properties concerning the stellar content (color at the effective radius, (B-R)$_{e}$, total extrapolated absolute magnitude, $M_{T}$, and star formation rate, SFR) of the H{\sc i}PASS sample along with similar histograms for the NFGS galaxies; the solid and dashed lines represent B-band and R-band data respectively; distance dependent quantities are computed with $h$=0.70.}
\label{fig9}
\end{figure}
\begin{figure}
\plotone{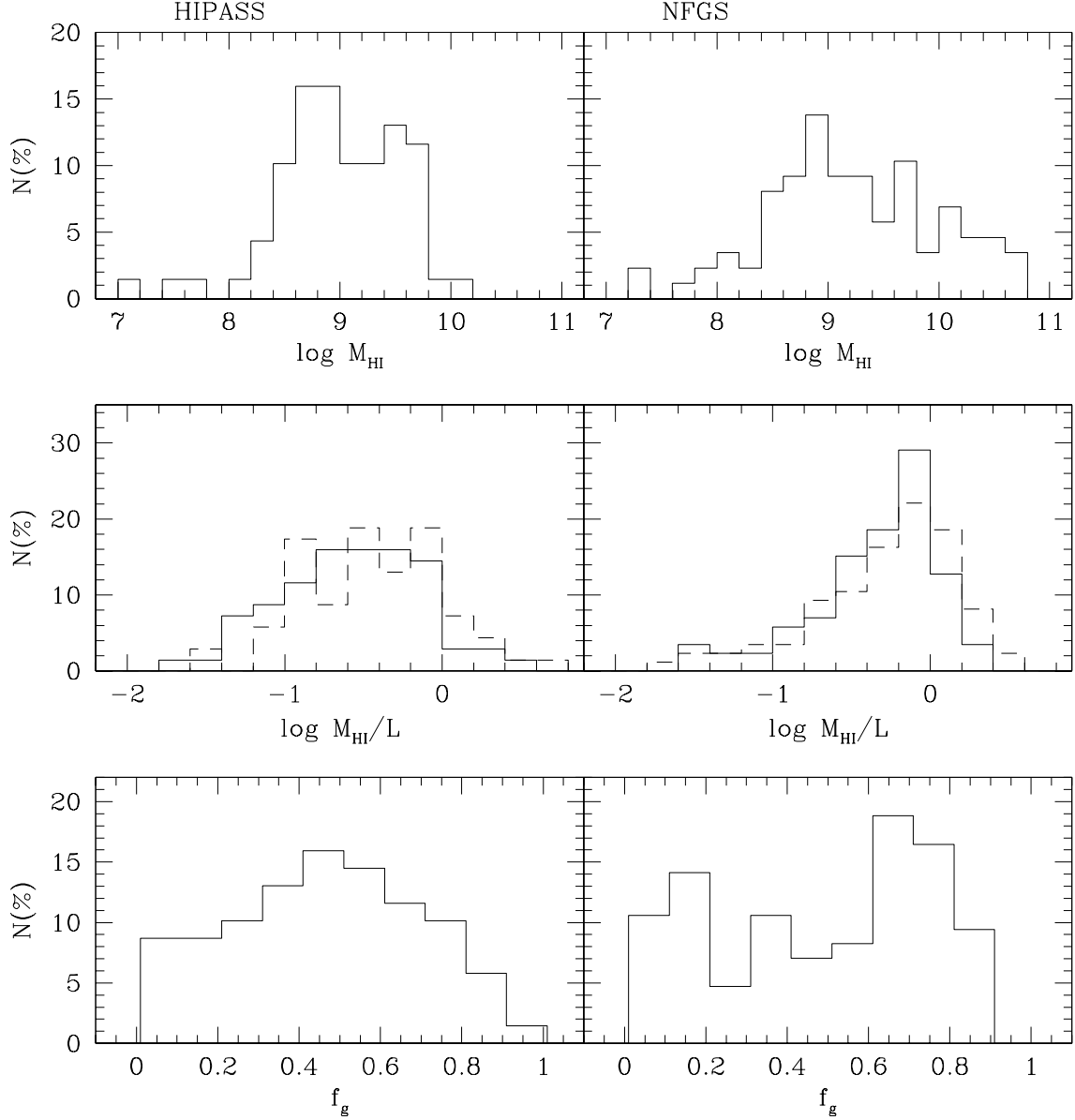}
\caption{The properties concerning the gas content (H{\sc i} mass, $M_{HI}$, gas-to-light ratio, $M_{HI}/L$, and gas-mass-fraction, $f_{g}$) of the H{\sc i}PASS sample along with similar histograms for the NFGS galaxies; the solid and dashed lines represent B-band and R-band data respectively; distance dependent quantities are computed with $h$=0.70.}
\label{fig10}
\end{figure}
\begin{figure}
\plotone{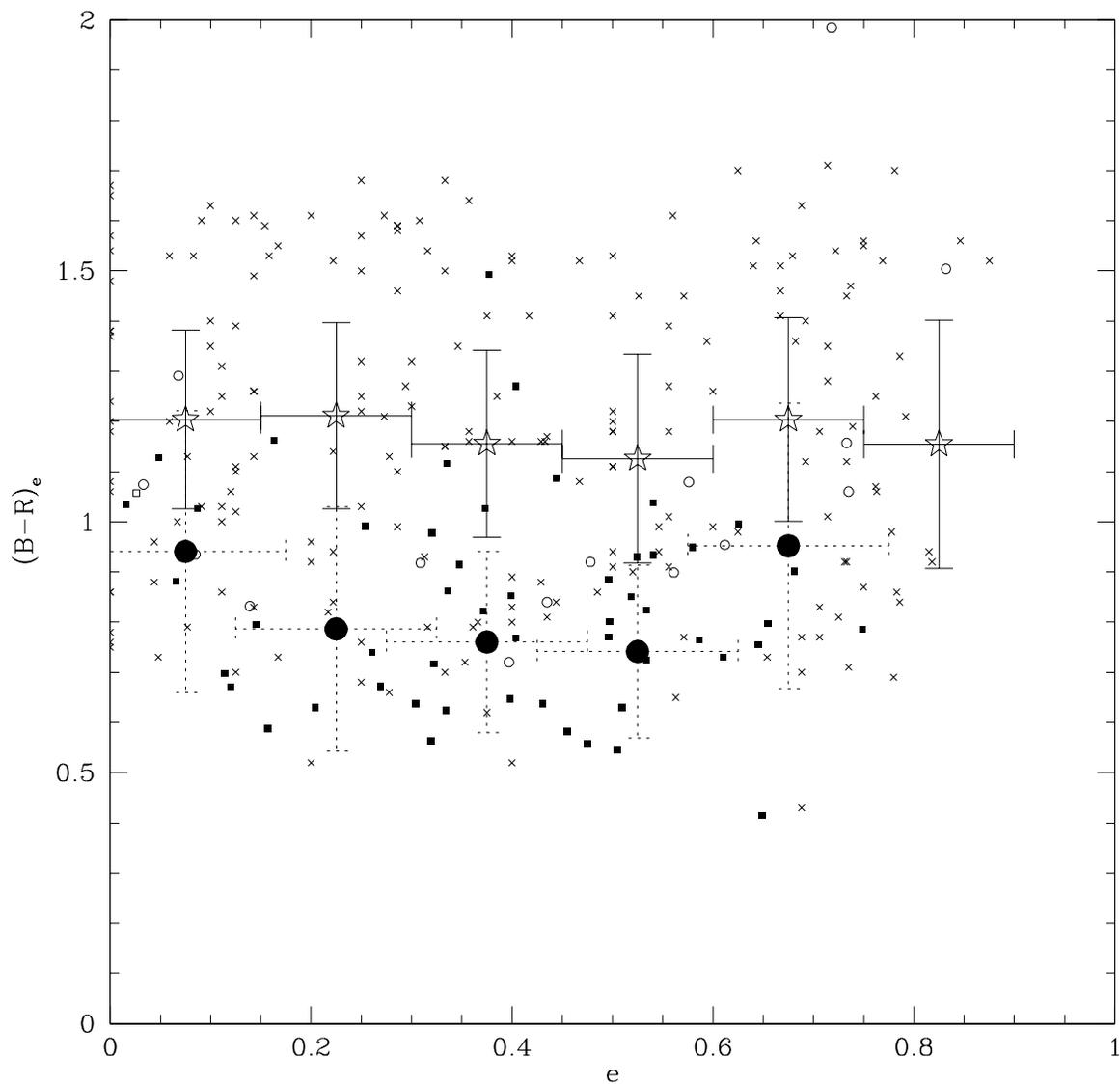}
\caption{Ellipticity (e=1-$\frac{b}{a}$) versus (B-R)$_{e}$ color.  NFGS galaxies are represented by $\times$'s; closed boxes represent H{\sc i}PASS galaxies; open circles represent H{\sc i}PASS galaxies for which the calibration is questionable; the open box is M83.  The mean values for (B-R)$_{e}$ for bins in ellipticity are plotted with the data; stars represent the NFGS galaxies; closed circles represent the H{\sc i}PASS galaxies.}
\label{fig11}
\end{figure}

\placetable{tab7}
\placefigure{fig8}
\placefigure{fig9}
\placefigure{fig10}
\placefigure{fig11}
The H{\sc i} and optical properties of the H{\sc i}PASS sample do not exactly match those of the dwarfs in the survey of \citet{scho01}.  Their sample exhibits a slightly different trend between gas-to-light ratio and central surface brightness and absolute magnitude than the spiral galaxies in the survey of \citet{dej96}.  The H{\sc i}PASS galaxies, however, continue along the same trends as the NFGS galaxies with a similar amount of scatter (see Fig. 12).\par

\begin{figure}
\plotone{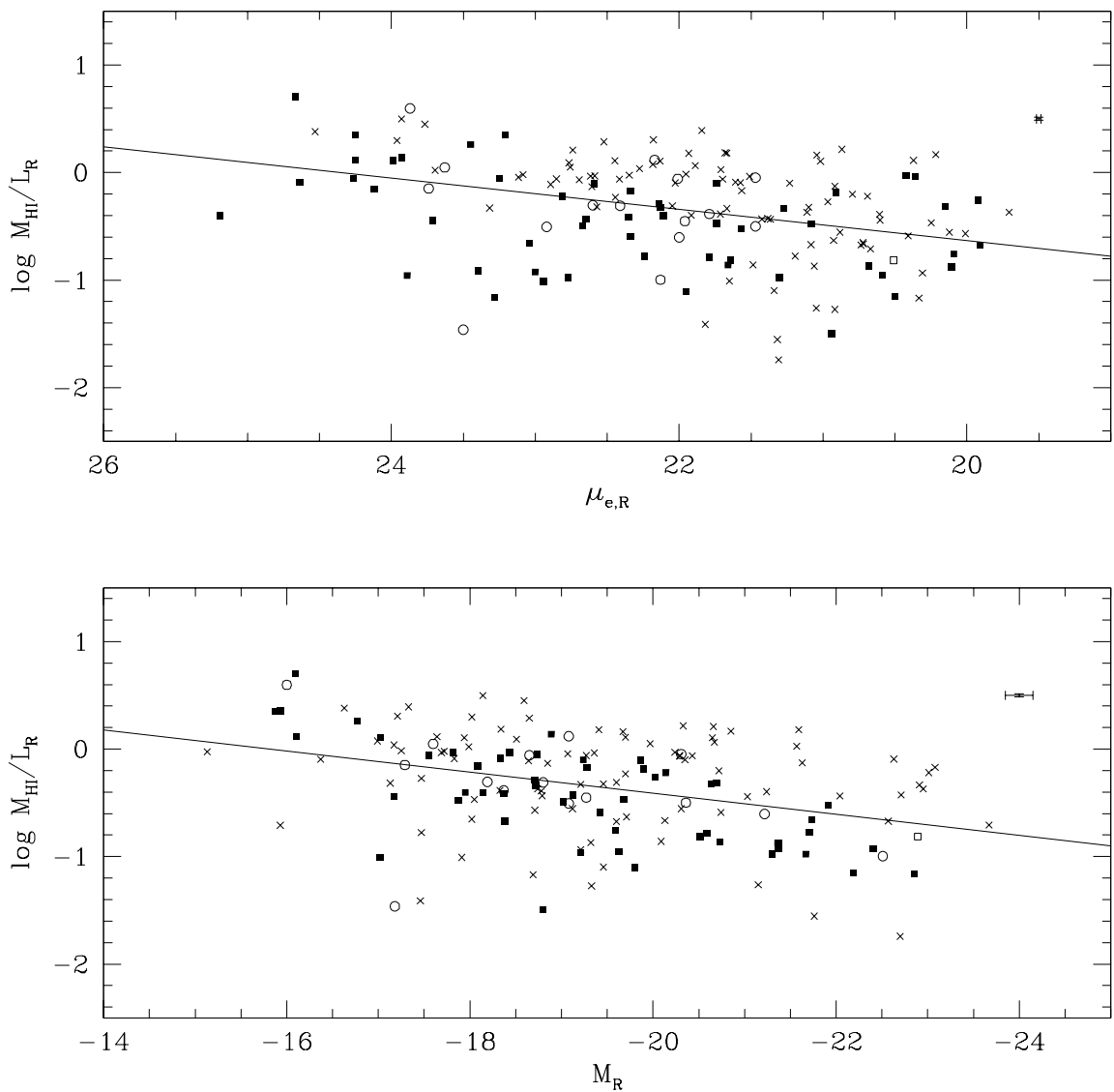}
\caption{The R-band gas-to-light ratio as a function of surface brightness at the effective radius and absolute magnitude. The solid lines are linear least-squares fits to the data.  NFGS galaxies are represented by $\times$'s; closed boxes represent H{\sc i}PASS galaxies; open circles represent H{\sc i}PASS galaxies for which the calibration is questionable; the open box is M83.  The displayed error bars are the median 1$\sigma$ errors for the H{\sc i}PASS galaxies.}
\label{fig12}
\end{figure}

\placefigure{fig12}

\subsection{Star Formation Properties}
For the calculation of the SFRs, the H$\alpha$ luminosities were determined using the corrected H$\alpha$ fluxes from the isophotal photometry measurements, the corrected radial velocities, and $h$=0.70.  From \citet{ken94}, the total SFR is given by\par
\begin{equation}
\mbox{SFR}_{total}=\frac{L_{H\alpha}}{1.26\times 10^{41} \mbox{ergs s}^{-1}} \mbox{ M}_{\odot} \mbox{ yr}^{-1}
\end{equation}
\noindent
SFRs were computed for the IRAS galaxies contained within the NFGS using the calibrated H$\alpha$ fluxes reported by \citet{kew02}, the corrected radial velocities reported by \citet{jan00}, and $h$=0.70.  These H$\alpha$ fluxes were corrected for extinction using the H$\alpha$ to H$\beta$ flux ratios measured by \citet{jan00}, the extinction curve of \citet{odo94}, and $R_{V}$=3.1 so that the distribution of these SFRs could be directly compared with the distribution of the SFRs computed for the H{\sc i}PASS galaxies.  It was found that most of the SFRs for the H{\sc i}PASS sample were $\sim$0.01 to 10 M$_{\odot}$ yr$^{-1}$ with a median value less than 1 M$_{\odot}$ yr$^{-1}$ (see Fig. 9).  This is typical for LSB and dwarf galaxies \citep{mih99}.  The NFGS sample has a higher fraction of galaxies with SFRs larger than 1 M$_{\odot}$ yr$^{-1}$ while missing many of the galaxies with SFRs$<$0.03 M$_{\odot}$ yr$^{-1}$ that appear in the H{\sc i}PASS sample.\par
The possible star formation histories for the galaxies can be explored by plotting their colors versus both their H$\alpha$ equivalent widths and their SFRs per unit H{\sc i} mass.  Fig. 13 demonstrates that the H{\sc i}PASS galaxies roughly follow the same trend between equivalent width and B-R color as the bluest NFGS galaxies with a slightly higher fraction of lower equivalent width galaxies.  Plotted with the data are model values generated with the PEGASE \citep{fio97} population synthesis code for three different star formation scenarios:
\begin{enumerate}
\item $\Psi$(t)=$\Psi_{\circ}$
\item $\Psi$(t)=$\Psi_{\circ} \cdot$exp[-t/$\tau$]
\item $\Psi$(t)=$\Psi_{\circ} \cdot f_{g}^{1.4}$ (Schmidt Law)
\end{enumerate}
where $\Psi$ is the SFR per unit galaxy mass (i.$\:$e. the combined mass of stars and gas), $\tau$ is an e-folding time scale, and $f_{g}$ is the gas mass fraction.  For all models, solar metallicity was used and the IMF applied was that used by \citet{ken83}.  For the exponentially decreasing SFR scenario, e-folding times of 1, 5, and 10 Gyr were used; the initial star formation rates per galaxy mass were set to $\tau^{-1}$ and 2$\cdot \tau^{-1}$.  For $\Psi_{\circ}=2 \cdot \tau^{-1}$, the star formation was forced to cease at 1.2, 1.8, and 8.9 Gyr for the three e-folding times used because the gas was completely consumed by these times.  For both the constant SFR and Schmidt Law scenarios, three different initial SFRs were used (referred to as low, medium, and high in Fig. 13 and 14).  For both scenarios, the initial SFRs were chosen to reproduce the observed range in SFR per unit H{\sc i} mass.\par
The H$\alpha$ equivalent widths produced by all of the models will most likely overestimate what is observed given that the extinction of nebular H$\alpha$ emission, $A_{H\alpha}$, tends to be higher than that for the stellar population at 6563 $\mbox{\AA}$, $A_{6563}$.  \citet{cal01} reports that for a typical spiral galaxy, the difference between these two values is about 0.5 mag; this implies that the observed equivalent widths will be lower by a factor of about 1.6.  To reproduce this effect within the models, the model H$\alpha$ equivalent widths were multiplied by a factor of 0.63.  Model curves for each of the models for ages spanning 5 to 20 Gyr are plotted with the data in Fig. 13 and 14.\par
Fig. 14 shows that there is no clear trend between SFR per unit H{\sc i} mass and color, but the range of values can be used to further check the models.  Model values for the gas-to-light ratios were computed assuming $M_{H}=0.7 \cdot M_{gas}$ and no molecular gas.  It is immediately apparent that the exponentially decreasing models with lower initial SFRs cannot explain what is observed; the higher initial SFR models need to be included for this scenario to be able to explain the redder galaxies with high values of SFR per unit H{\sc i} mass.  Again, the constant SFR models cannot explain the redder galaxies; the Schmidt Law models are able to reproduce the observed range in parameter space inhabited by the galaxies.  As may be expected, all the models indicate that there is a rough correlation between color and mean age with redder galaxies being older than bluer ones.\par
Given that the observed trend between the mean SFR surface density and mean gas column density in star forming galaxies that the Schmidt Law models are based on has been observed to hold over five to six orders of magnitude \citep{ken98} and that they are able to reproduce the observed typical H$\alpha$ equivalent width and range of SFRs per unit H{\sc i} mass at any given color, the Schmidt Law models were used to estimate a mean age for the stellar population in each galaxy.  Two additional Schmidt Law models were run to more fully cover the observed parameter space for color and SFR per unit H{\sc i} mass, one with an initial SFR in between the low and medium initial SFR models and one with an initial SFR in between the medium and high initial SFR models for ages from 5 to 20 Gyr.\par
To explore the effect of metallicity on the ages computed with these models, all five were run with five different metallicities, Z=0.002, 0.004, 0.01, 0.02, and 0.04.  The resulting five grids of models, one for each value of Z, were interpolated to obtain a mean age for each galaxy at each metallicity using the galaxy's color and SFR per unit H{\sc i} mass.  As a check, a model prediction for the H$\alpha$ equivalent width was obtained from each interpolated model and multiplied by 0.63 as described above.  These were compared to the observed values, and it was found that the agreement between the observed and predicted values was similar for all metallicities; for both samples, the mean deviation was $\sim$0 and the rms deviation was $\sim$0.4 dex.  The fact that the average deviation was nearly zero implies that the model ages on average provide a good prediction of the H$\alpha$ equivalent width for a given metallicity.  However, the fact that the model values were typically higher or lower than what was observed by a factor of about 2.5 implies that no one metallicity can explain all of the data well for each sample.\par
The median ages for the H{\sc i}PASS galaxies using the five model grids were 5.2, 4.7, 3.6, 2.9, and 2.5 Gyr respectively; the median values for the NFGS galaxies that had both measured SFRs and H{\sc i} masses were 8.6, 7.6, 7.1, 5.8, and 4.5 Gyr.  For both samples, the rms deviation about the median values for the calculated ages were roughly constant from one model grid to the next.  For the H{\sc i}PASS sample, the deviation was about 3.2 Gyr, and the rms deviation was about 4.5 Gyr for the NFGS galaxies.  From these results, it can be seen that the difference in color and SFR per unit H{\sc i} mass between the two samples can be explained solely by a difference in metallicity only if the typical abundance is between $\frac{1}{10}$ and $\frac{1}{5}$Z$_{\odot}$ for the H{\sc i}PASS sample and between 1 and 2Z$_{\odot}$ for the NFGS galaxies.  This particular scenario is unlikely given that even though LSB galaxies have been observed to have lower gas phase abundances \citep{mcg94}, there is no observational evidence that the typical metallicity for local LSB galaxies is as lower than $\frac{1}{5}$Z$_{\odot}$.  Some difference in typical mean age must then be invoked to explain the discrepancy in color.\par
It should be noted, however, that no attempt was made to account for the effect of internal reddening on the (B-R) colors; such reddening would cause the ages to be overestimated.  For example, including a correction for an extra 0.5 mag of extinction in the B-band (E(B-V)$\approx$0.12) for each of the H{\sc i}PASS galaxies reduces the median ages obtained from the five model grids by about 1 to 2 Gyr.  It should also be noted that conclusions based on these results hinge on the assumption that all the galaxies have the same IMF that is constant with time.  Overall, these results imply that on average, the H{\sc i}PASS galaxies have formed the bulk of their stars more recently than the redder NFGS galaxies but that the star formation histories for most of the H{\sc i}PASS galaxies are quite similar to the bluest NFGS galaxies.  However, there is a small group of blue galaxies within the H{\sc i}PASS sample with H$\alpha$ equivalent widths that are lower than is expected (log W$_{H\alpha}$ less than $\sim$1) for their colors that does not appear within the NFGS.\par
\begin{figure}
\plotone{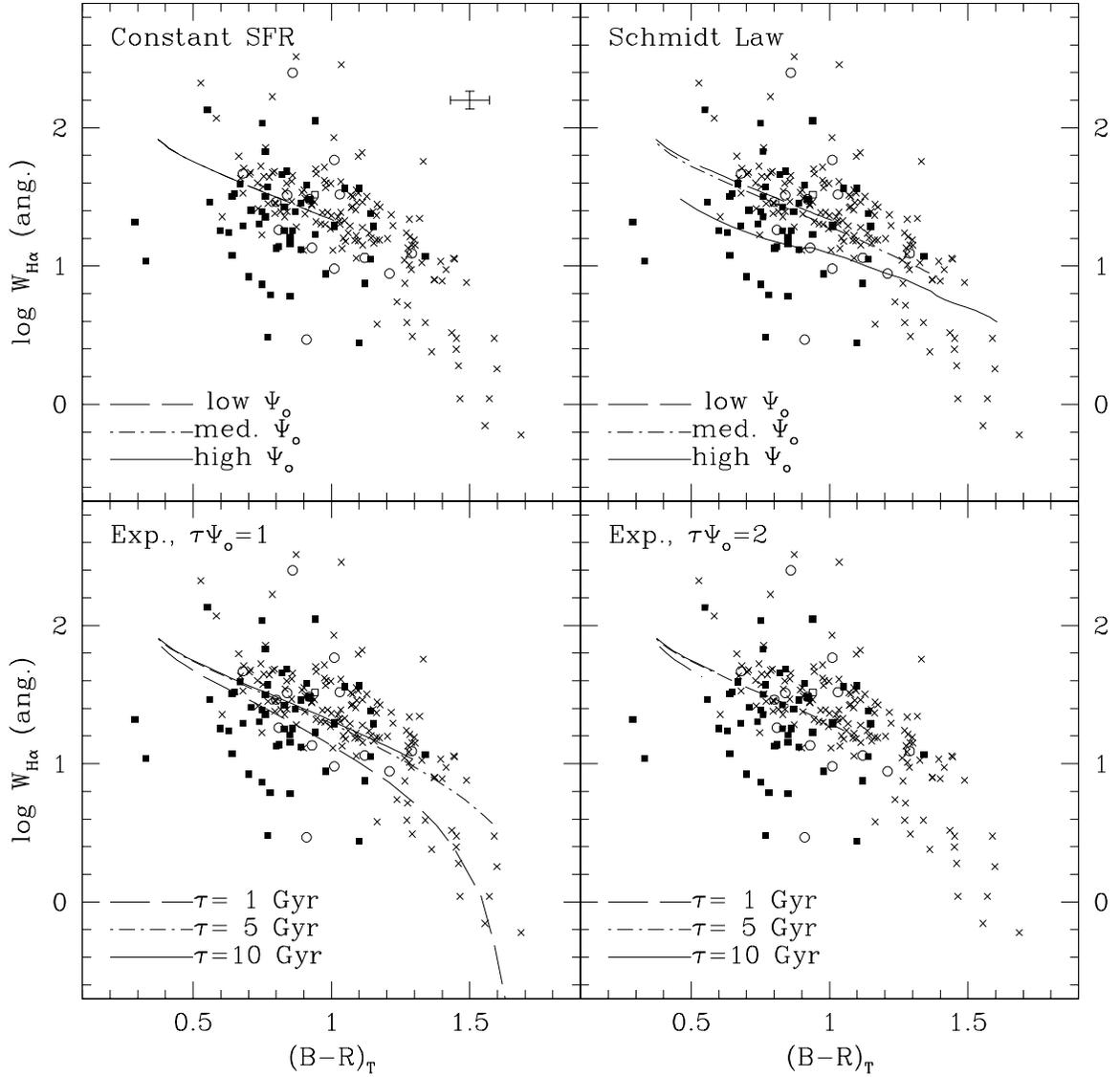}
\caption{Reddening corrected, total extrapolated (B-R)$_{T}$ color versus H$\alpha$ equivalent width with model predictions based on different star formation histories.  Here, $\Psi_{\circ}$ refers to the initial SFR in units of M$_{gal}$ Myr$^{-1}$ where M$_{gal}$ is the combined mass of stars and gas; $\tau$ refers to the e-folding time for the exponential models (i.$\:$e. $\tau \Psi_{\circ}$ gives the total amount of mas formed at t$\rightarrow \infty$).  See Sec. 4.3 for a detailed discussion of the model curves plotted.  NFGS galaxies are represented by $\times$'s; closed boxes represent H{\sc i}PASS galaxies; open circles represent H{\sc i}PASS galaxies for which the calibration is questionable; the open box is M83.  The displayed error bars are the median 1$\sigma$ errors for the H{\sc i}PASS galaxies.}
\label{fig13}
\end{figure}
\begin{figure}
\plotone{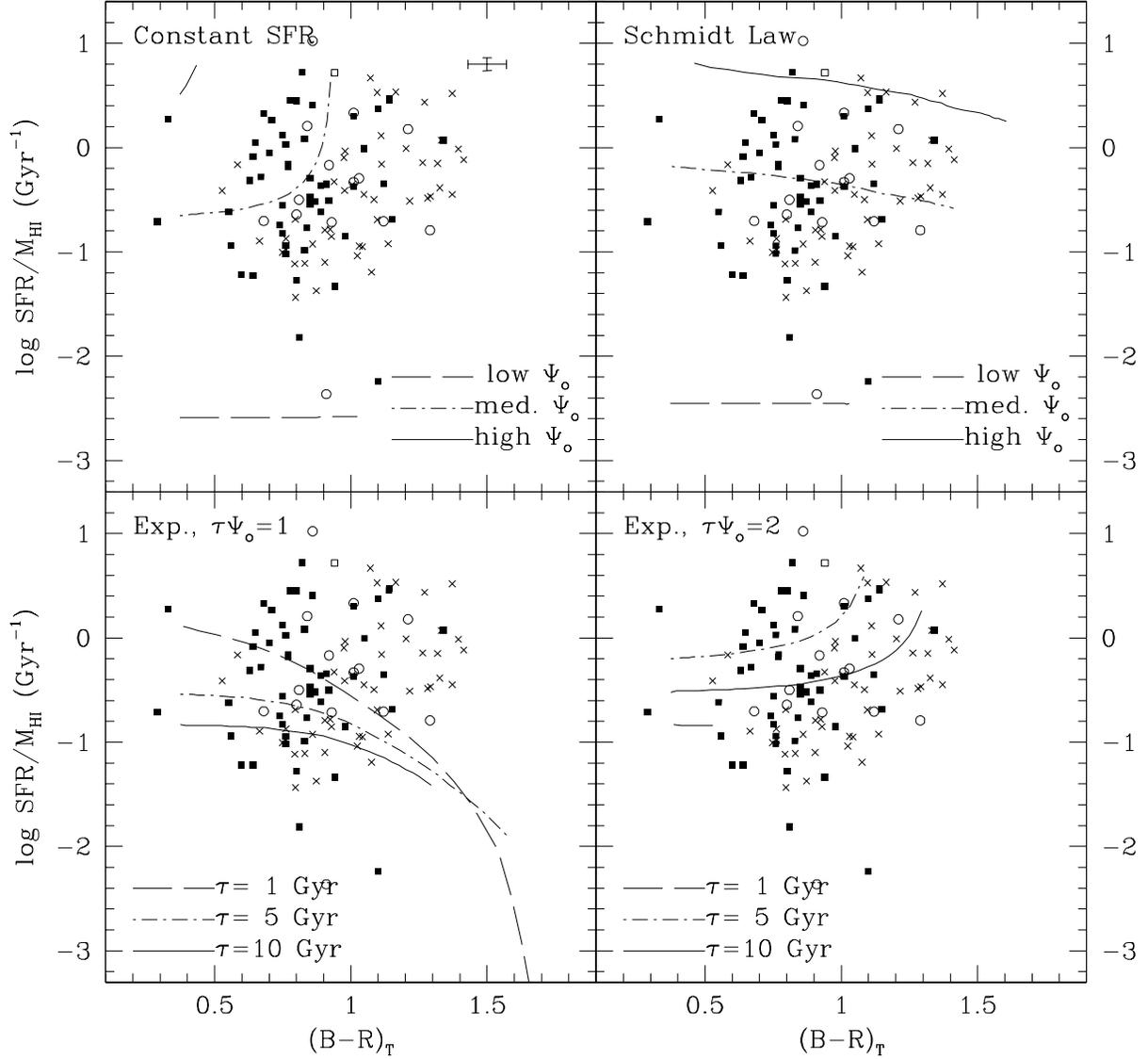}
\caption{Reddening corrected (B-R)$_{T}$ color versus SFR per unit HI mass.  The lines are generated by the same models as the ones plotted in Fig. 13.  NFGS galaxies are represented by $\times$'s; closed boxes represent H{\sc i}PASS galaxies; open circles represent H{\sc i}PASS galaxies for which the calibration is questionable; the open box is M83.  The displayed error bars are the median 1$\sigma$ errors for the H{\sc i}PASS galaxies.}
\label{fig14}
\end{figure}

\placefigure{fig13}
\placefigure{fig14}
The PEGASE models were also used in the calculation of the gas mass fractions for the galaxies.  All of the models run produce values that follow a similar trend between B-R color and stellar mass-to-light ratio, $\Upsilon_{\ast}$.  The trend is well approximated by log$(\Upsilon_{\ast_{B}})$=1.41(B-R)-1.55.  The distributions of gas mass fractions for the samples are plotted in Fig. 10.\par

\subsection{Surface Brightness and Number, Luminosity, and SFR Densities}
The effect of any surface brightness bias on the measured value of the number, luminosity, and SFR densities can be explored by computing the values as functions of surface brightness for the H{\sc i}PASS sample.  All but four of the H{\sc i}PASS galaxies have H{\sc i} masses that would make them detectable within H{\sc i}PASS at or beyond our radial velocity limit of 2500 km s$^{-1}$ for $h$=0.70.  Of these four, two are below the 3$\sigma$ detection limit of 40 mJy; the four galaxies have a wide range of surface brightness values.  Therefore, we have taken the H{\sc i}PASS sample to be volume limited for calculating the number, luminosity, and SFR densities as functions of surface brightness while including a correction factor of $\frac{132}{69}$ to account for the galaxies that met the selection criteria but were not imaged.\par
Given the somewhat complex selection criteria used for the NFGS, computing these densities as functions of $\mu_{e}$ for that sample would be difficult.  The larger volume covered by the NFGS would also reduce the usefulness of any direct comparison of these quantities between the NFGS and the H{\sc i}PASS sample.  We therefore asses the degree to which the surface brightness bias may effect the measured number, luminosity, and SFR densities by exploiting the difference in the surface brightness coverage between these two samples.  To do this, we compute the mean and standard deviation for the distributions for $\mu_{e}$ for both the entire NFGS and for the IRAS galaxies contained within the NFGS for which SFRs have been measured.  The mean $\pm$2$\sigma$ for the the NFGS are displayed in Fig. 15 along with $\mu_{e}$ versus number and luminosity densities for the H{\sc i}PASS sample; similar values for the NFGS/IRAS galaxies are displayed with the H{\sc i}PASS sample values for  $\mu_{e}$ versus SFR density.  For the three $\mu_{e}$ bins that are more than 2$\sigma$ fainter than the mean for the NFGS, the total number and luminosity density was calculated and compared with the total values for all the bins.  It was found that 14$\pm$5.0(1$\sigma$)\% of the total number density is contained within these three bins.  The bins contain a much smaller fraction of the total luminosity density; the value ranges from 0.5$\pm$0.2(1$\sigma$)\% if the isophotal fluxes are used and 0.8$\pm$0.3(1$\sigma$)\% if the total extrapolated fluxes are used.  Four surface brightness bins are more than 2$\sigma$ fainter than the mean for the NFGS/IRAS galaxies.  It was found that 3.9$\pm$1.6(1$\sigma$)\% of the total SFR density is contained within these bins.\par
\begin{figure}
\plotone{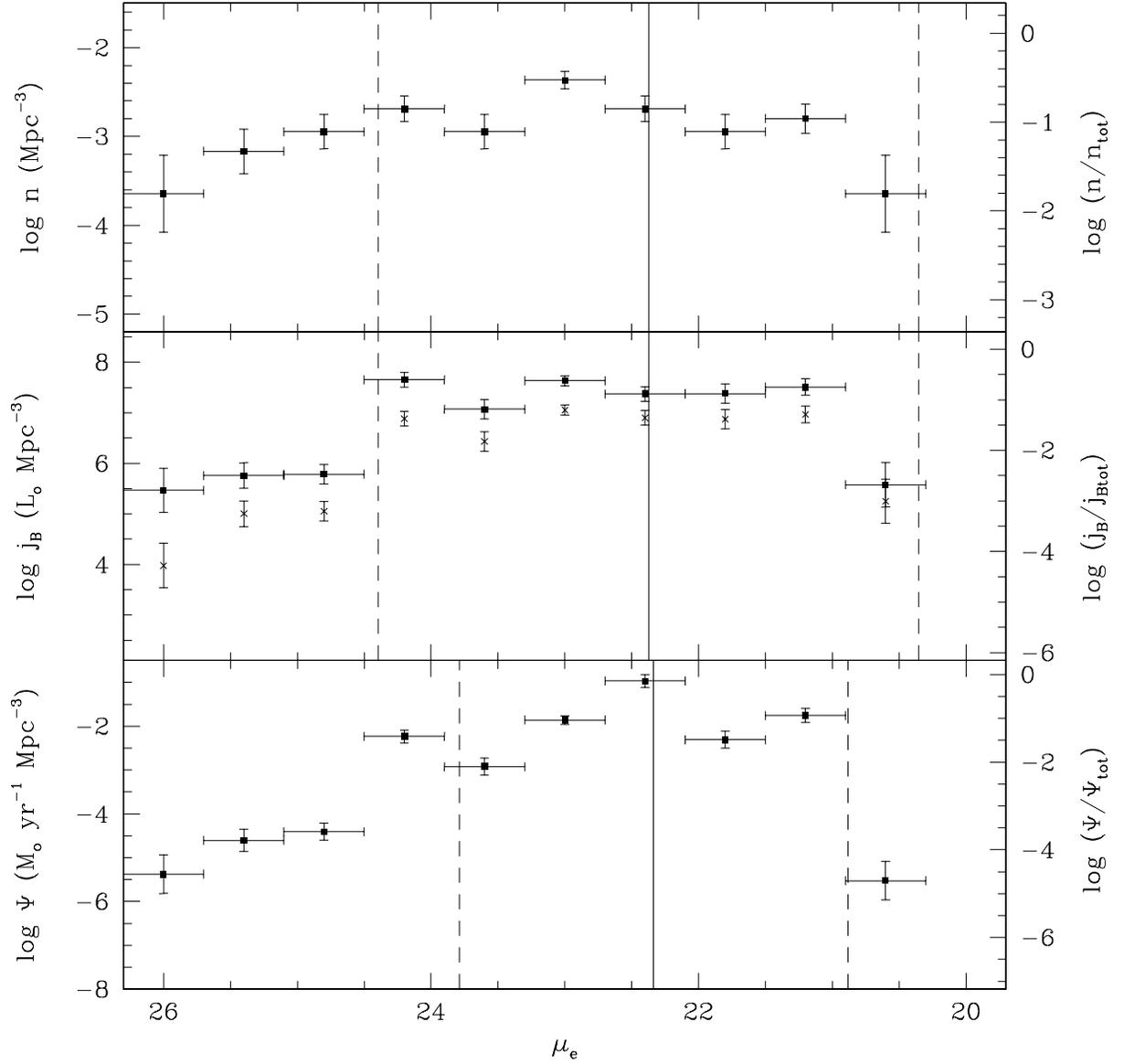}
\caption{Half-light surface brightness versus number (top panel), luminosity (middle panel), and SFR (bottom panel) densities for the H{\sc i}PASS sample; the vertical error bars are 1$\sigma$ error bars; the horizontal bars represent the width of the surface brightness bins.  The $\times$'s in the middle panel represent values calculated with isophotal fluxes; the boxes represent values calculated with total extrapolated fluxes.  The solid vertical lines in the top and middle panels represent the mean value for $\mu_{e}$ for the NFGS galaxies; the dashed lines represent the mean $\pm 2 \sigma$; the lines in the lower panel represent the same quantities for the NFGS/IRAS galaxies.}
\label{fig15}
\end{figure}

\placefigure{fig15}

\section{Conclusions}
\subsection{The Surface Brightness Bias}
The existence of a bias toward higher surface brightness disks in flux limited optical catalogs is clearly illustrated by the comparison of properties of the H{\sc i}PASS and NFGS samples.  Nearly 25\% of the galaxies in the H{\sc i}PASS sample, which was chosen from the H{\sc i}PASS catalog purely by declination and radial velocity, have half-light surface brightness values $\geq$24.  Only about 3.6\% of the optically selected NFGS galaxies have values that lie beyond this limit; this value increases slightly to 4.4\% if only spiral galaxies are considered.  The existence of this bias has been known and examined for quite some time \citep[e.$\:$g.][]{dis76,imp97,bot97}.  However, the degree to which this bias effects our knowledge of the properties of the local galaxy population is largely dependent on the properties of interest.\par

\subsection{The Galaxy Luminosity Function}

\citet{zwa01} have demonstrated that for a sample of galaxies similar to the H{\sc i}PASS sample taken from the Arecibo H{\sc i} Strip Survey (AH{\sc i}SS), the luminosity function is very similar to those found for optical samples.  Other authors \citep[e.$\:$g.][]{bro01} have confirmed that the inclusion of LSB disks in optically selected samples changes the inferred luminosity function very little.  Our results are in agreement with this conclusion as Fig. 15 implies that a sample biased toward higher surface brightness galaxies will miss less than 1\% of the total local luminosity density.\par
However, it should be noted that Fig. 15 also indicates that the degree to which LSB galaxies contribute to the local galaxy LF depends greatly on how their luminosities are measured.  Given that the central surface brightness values for these objects are quite low, it is easy to miss a significant portion of the light emitted by these objects.  For example, for a galaxy with $\mu_{B, \circ}$=24, if the galaxy is approximated with a pure exponential disk, the total flux can be up to a factor of three times larger than the isophotal flux depending on the scale length of the disk.  This is seen most prominently in the lowest surface brightness bin in the middle panel of Fig. 15; for this bin, the luminosity density calculated using the total extrapolated fluxes is about 30 times greater than that calculated using the isophotal luminosities.\par

\subsection{Star Formation in the Local Universe}

Fig. 15 indicates that LSB galaxies can contribute a small amount to the total local star formation that may be missed by optical surveys but that also appears to be overlooked by IR surveys such as the one performed by IRAS.  This is mainly caused by the moderate sensitivity of the IRAS survey ($\sim$0.6 Jy at 60 $\mu$m \citep{sau95}), but still reflects the fact that the optimal wavelength regime in which to search for these objects is most likely the radio at rest frame 21 cm.  Based on Fig. 15, the exclusion of these galaxies from samples selected in the optical (or IR) used to calculate the local SFR may miss up to 3 to 4\% of the total amount of star formation.\par
It is also clear that the H{\sc i}PASS sample contains a higher fraction of bluer galaxies that have high H$\alpha$ equivalent widths, indicating that the current star formation in these galaxies is fairly high when compared to how many stars have formed in them in the past.  Fig. 13 and 14 demonstrate that the colors, equivalent widths, and SFRs per unit H{\sc i} mass are what is expected for stellar populations with young mean ages; Schmidt Law models estimate the typical mean age to be about 4.0 Gyr for Z=$\frac{1}{3}$Z$_{\odot}$.  This is nearly 2 Gyr smaller than the median age estimated for the NFGS for solar metallicity, implying that on average, bluer more LSB galaxies that are present in lower numbers in the NFGS formed the majority of their stars more recently.  The range in ages for the H{\sc i}PASS sample is $\sim$1 Gyr smaller than that for the NFGS, implying that the majority of the star formation activity that has occurred in similar objects has taken place over a smaller range in redshift space.  If it is assumed that the typical mean age for the H{\sc i}PASS sample is about 4 Gyr with an rms deviation of 3.2 Gyr, then one would expect the progenitors of the blue LSB galaxies seen locally to contain few if any stars at redshifts of $\sim$1 or greater (i.$\:$e. look-back time greater than about 7.2 Gyr).\par
The authors would like to thank the NOAO TAC for allocation of observing time and the CTIO staff for expert assistance at the telescope as well as the referee for useful comments and suggestions.

\clearpage

\clearpage

\clearpage

\clearpage

\clearpage

\clearpage

\clearpage

\clearpage

\clearpage

\clearpage

\clearpage

\clearpage

\clearpage

\clearpage

\clearpage

\clearpage

\input{tab1.tex}

\clearpage

\clearpage

\input{tab2.tex}

\clearpage

\input{tab3.tex}

\clearpage

\input{tab4.tex}

\clearpage

\input{tab5.tex}

\clearpage

\input{tab6.tex}

\clearpage

\input{tab7.tex}

\end{document}

%% file: tab1.tex
\begin{deluxetable}{lrrrrrrr}
\tablecolumns{8}
\tablewidth{0pc}
\tablecaption{Observations}
\tablehead{
\colhead{Galaxy} & \colhead{UT Date} & \colhead{RA\tablenotemark{a}} & \colhead{DEC\tablenotemark{b}} & 
\colhead{$\tau_{B}$\tablenotemark{c}} & \colhead{$\tau_{R}$} & \colhead{$\tau_{H\alpha}$} & \colhead{H$\alpha$ Filter\tablenotemark{d}} \\
\colhead{} & \colhead{} & \colhead{(2000.0)} & \colhead{(2000.0)} & \colhead{(s)} & \colhead{(s)} & \colhead{(s)} & \colhead{}}

\startdata
\sidehead{CTIO March 2-9, 2000}
\hline
        NGC1313 & 000303 & 03 18 16.48 & -66 30 33.8 & 2$\times$300 & 2$\times$300 & 2$\times$1200 & 657 nm \\ 
    ESO004-G017 & 000303 & 05 04 29.36 & -87 01 45.0 & 2$\times$300 & 2$\times$300 & 2$\times$1200 & 660 nm \\ 
    ESO035-G020 & 000303 & 08 03 18.61 & -77 04 18.7 & 2$\times$300 & 2$\times$300 & 2$\times$1200 & 660 nm \\ 
  HIPASS1039-71 & 000303 & 10 39 40.00 & -71 55 23.0 & 2$\times$300 & 2$\times$300 & 2$\times$1200 & 660 nm \\ 
    ESO092-G006 & 000303 & 10 03 18.96 & -64 58 03.9 & 2$\times$300 & 2$\times$300 & 2$\times$1200 & 661 nm \\ 
         IC4662 & 000303 & 17 47 06.38 & -64 38 25.1 & 2$\times$300 & 2$\times$300 & 2$\times$1200 & 657 nm \\ 
    ESO084-G040 & 000304 & 04 44 59.88 & -62 42 24.5 & 2$\times$300 & 2$\times$300 & 2$\times$1200 & 660 nm \\ 
    ESO059-G001 & 000304 & 07 31 18.20 & -68 11 16.8 & 2$\times$300 & 2$\times$300 & 2$\times$1200 & 657 nm \\ 
    ESO060-G019 & 000304 & 08 57 26.20 & -69 03 35.9 & 2$\times$300 & 2$\times$300 & 2$\times$1200 & 660 nm \\ 
    ESO091-G007 & 000304 & 09 17 30.88 & -62 53 03.3 & 2$\times$300 & 2$\times$300 & 2$\times$1200 & 661 nm \\ 
    ESO092-G021 & 000304 & 10 21 05.37 & -66 29 27.2 & 2$\times$300 & 2$\times$300 & 2$\times$1200 & 661 nm \\ 
       NGC6438A & 000304 & 18 22 43.24 & -85 24 14.6 & 2$\times$300 & 2$\times$300 & 2$\times$1200 & 661 nm \\ 
    ESO086-G060 & 000305 & 06 08 09.26 & -63 35 16.8 & 2$\times$300 & 2$\times$300 & 2$\times$1200 & 660 nm \\ 
    ESO036-G006 & 000305 & 08 38 47.09 & -75 09 23.4 & 2$\times$300 & 2$\times$300 & 2$\times$1200 & 6600 $\mbox{\AA}$ \\ 
       NGC2397A & 000305 & 07 21 07.82 & -69 06 59.7 & 2$\times$300 & 2$\times$300 & 2$\times$1200 & 660 nm \\ 
        NGC2915 & 000305 & 09 26 11.49 & -76 37 35.6 & 2$\times$300 & 2$\times$300 & 2$\times$1200 & 657 nm \\ 
         IC3104 & 000305 & 12 18 46.06 & -79 43 33.8 & 2$\times$300 & 2$\times$300 & 2$\times$1200 & 657 nm \\ 
    ESO021-G003 & 000305 & 13 32 27.28 & -80 25 56.5 & 2$\times$300 & 2$\times$300 & 2$\times$1200 & 661 nm \\
    ESO085-G014 & 000306 & 04 54 42.75 & -62 47 59.1 & 2$\times$300 & 2$\times$300 & 2$\times$1200 & 660 nm \\ 
        NGC1892 & 000306 & 05 17 07.98 & -64 57 41.0 & 2$\times$300 & 2$\times$300 & 2$\times$1200 & 660 nm \\ 
        NGC2836 & 000306 & 09 13 44.60 & -69 20 05.1 & 2$\times$300 & 2$\times$300 & 2$\times$1200 & 660 nm \\ 
    ESO035-G021 & 000306 & 08 09 55.49 & -74 30 41.8 & 2$\times$300 & 2$\times$300 & 2$\times$1200 & 660 nm \\ 
    ESO019-G004 & 000306 & 10 48 50.14 & -80 14 12.4 & 2$\times$300 & 2$\times$300 & 2$\times$1200 & 660 nm \\ 
        NGC6300 & 000306 & 17 16 59.22 & -62 49 11.2 & 2$\times$300 & 2$\times$300 & 2$\times$1200 & 6600 $\mbox{\AA}$ \\ 
    ESO085-G047 & 000307 & 05 07 43.86 & -62 59 24.3 & 2$\times$300 & 2$\times$300 & 2$\times$1200 & 660 nm \\ 
    ESO037-G004 & 000307 & 09 32 37.51 & -74 15 16.0 & 2$\times$300 & 2$\times$300 & 2$\times$1200 & 660 nm \\ 
    ESO090-G004 & 000307 & 08 38 36.86 & -64 20 32.4 & 2$\times$300 & 2$\times$300 & 2$\times$1200 & 660 nm \\ 
    ESO061-G017 & 000307 & 09 52 33.70 & -69 04 04.2 & 2$\times$300 & 2$\times$300 & 2$\times$1200 & 660 nm \\ 
       NGC3136A & 000307 & 10 03 33.49 & -67 26 52.5 & 2$\times$300 & 2$\times$300 & 2$\times$1200 & 661 nm \\ 
    ESO038-G011 & 000307 & 11 20 58.10 & -75 52 45.4 & 2$\times$300 & 2$\times$300 & 2$\times$1200 & 660 nm \\ 
    ESO140-G019 & 000307 & 18 22 46.46 & -62 16 12.8 & 2$\times$300 & 2$\times$300 & 2$\times$1200 & 657 nm \\ 
  HIPASS0635-70 & 000308 & 06 35 36.00 & -70 52 55.0 & 2$\times$300 & 2$\times$300 & 2$\times$1200 & 660 nm \\ 
        NGC2442 & 000308 & 07 36 23.77 & -69 31 49.5 & 2$\times$300 & 2$\times$300 & 2$\times$1200 & 660 nm \\ 
       NGC2788B & 000308 & 09 03 37.19 & -67 57 58.9 & 2$\times$300 & 2$\times$300 & 2$\times$1200 & 660 nm \\ 
    ESO060-G007 & 000308 & 08 27 26.50 & -71 04 19.3 & 2$\times$300 & 2$\times$300 & 2$\times$1200 & 660 nm \\ 
    ESO037-G015 & 000308 & 10 25 41.99 & -76 30 18.3 & 2$\times$300 & 2$\times$300 & 2$\times$1200 & 660 nm \\ 
        NGC5068 & 000308 & 13 18 55.24 & -21 02 21.5 & 2$\times$300 & 2$\times$300 & 2$\times$1200 & 657 nm \\ 
         IC4710 & 000308 & 18 28 38.16 & -66 58 54.3 & 2$\times$300 & 2$\times$300 & 2$\times$1200 & 657 nm \\ 
  HIPASS0653-73 & 000309 & 06 53 50.00 & -73 40 35.0 & 2$\times$300 & 2$\times$300 & 2$\times$1200 & 6600 $\mbox{\AA}$ \\ 
   ESO060-IG003 & 000309 & 08 16 33.69 & -71 51 35.0 & 2$\times$300 & 2$\times$300 & 2$\times$1200 & 660 nm \\ 
        NGC3059 & 000309 & 09 50 07.95 & -73 55 17.3 & 2$\times$300 & 2$\times$300 & 2$\times$1200 & 6600 $\mbox{\AA}$ \\ 
    ESO037-G010 & 000309 & 10 04 16.71 & -75 28 43.0 & 2$\times$300 & 2$\times$300 & 2$\times$1200 & 660 nm \\ 
            M83 & 000309 & 13 37 00.23 & -29 52 04.5 & 2$\times$300 & 2$\times$300 & 2$\times$1200 & 657 nm \\ 
    ESO104-G022 & 000309 & 18 55 41.24 & -64 48 39.2 & 2$\times$300 & 2$\times$300 & 2$\times$1200 & 657 nm \\ 
\hline
\sidehead{CTIO October 23-28, 2000}
\hline
        NGC7098 & 001025 & 21 44 16.49 & -75 06 44.3 & 2$\times$480 & 2$\times$300 & 2$\times$1800 & 6600 $\mbox{\AA}$ \\ 
        NGC0802 & 001025 & 01 59 06.98 & -67 52 10.4 & 2$\times$480 & 2$\times$300 & 2$\times$1800 & 6600 $\mbox{\AA}$ \\ 
    ESO054-G021 & 001025 & 03 49 50.18 & -71 38 07.1 & 2$\times$420 & 2$\times$300 & 2$\times$1800 & 6600 $\mbox{\AA}$ \\ 
         IC2051 & 001025 & 03 52 02.29 & -83 49 56.6 & 2$\times$420 & 2$\times$300 & 2$\times$1800 & 6600 $\mbox{\AA}$ \\ 
    ESO027-G001 & 001026 & 21 52 27.81 & -81 31 50.5 & 2$\times$480 & 2$\times$300 & 2$\times$1800 & 6600 $\mbox{\AA}$ \\ 
    ESO013-G016 & 001026 & 01 32 48.37 & -79 28 26.4 & 2$\times$480 & 2$\times$300 & 2$\times$1800 & 6600 $\mbox{\AA}$ \\ 
        NGC1511 & 001026 & 03 59 35.73 & -67 38 06.6 & 2$\times$420 & 2$\times$300 & 2$\times$1800 & 6600 $\mbox{\AA}$ \\ 
    ESO119-G016 & 001026 & 04 51 29.19 & -61 39 03.4 & 2$\times$480 & 2$\times$300 & 2$\times$1800 & 6563 $\mbox{\AA}$ \\ 
    ESO027-G021 & 001027 & 23 04 19.50 & -79 28 01.1 & 2$\times$420 & 2$\times$300 & 2$\times$1800 & 6600 $\mbox{\AA}$ \\ 
        NGC0406 & 001027 & 01 07 24.12 & -69 52 35.3 & 2$\times$420 & 2$\times$300 & 2$\times$1800 & 6600 $\mbox{\AA}$ \\ 
        NGC1559 & 001027 & 04 17 37.29 & -62 47 03.6 & 2$\times$420 & 2$\times$300 & 2$\times$1800 & 6600 $\mbox{\AA}$ \\ 
    ESO085-G030 & 001027 & 05 01 30.02 & -63 17 33.9 & 2$\times$420 & 2$\times$300 & 2$\times$1800 & 6600 $\mbox{\AA}$ \\ 
        NGC1809 & 001027 & 05 02 05.63 & -69 34 07.9 & 2$\times$480 & 2$\times$300 & 2$\times$1200 & 6600 $\mbox{\AA}$ \\ 
         IC4870 & 001028 & 19 37 38.25 & -65 48 45.3 & 2$\times$420 & 2$\times$300 & 2$\times$1800 & 6600 $\mbox{\AA}$ \\ 
        NGC7661 & 001028 & 23 27 14.51 & -65 16 12.6 & 2$\times$480 & 2$\times$300 & 2$\times$1800 & 6600 $\mbox{\AA}$ \\ 
        NGC2082 & 001028 & 05 41 50.51 & -64 18 00.9 & 2$\times$480 & 2$\times$300 & 2$\times$1800 & 6600 $\mbox{\AA}$ \\ 
    ESO017-G002 & 001028 & 07 32 20.37 & -77 55 07.5 & 2$\times$420 & 2$\times$300 & 2$\times$1800 & 6600 $\mbox{\AA}$ \\ 
    ESO035-G009 & 001028 & 07 28 40.95 & -75 03 14.9 & 2$\times$600 & 2$\times$600 & $\cdots$ & $\cdots$ \\
\hline
\sidehead{CTIO November 2, 5, \& 6, 2000}
\hline
         IC5028 & 001102 & 20 43 21.86 & -65 38 47.7 & 2$\times$600 & 2$\times$420 & 2$\times$1200 & 660 nm \\ 
    ESO079-G005 & 001102 & 00 40 43.53 & -63 26 30.7 & 2$\times$480 & 2$\times$300 & 2$\times$1200 & 660 nm \\ 
    ESO080-G006 & 001102 & 01 47 16.94 & -62 58 13.3 & 2$\times$480 & 2$\times$300 & 2$\times$1200 & 660 nm \\ 
    ESO035-G018 & 001102 & 07 55 03.97 & -76 24 41.3 & 2$\times$480 & 2$\times$300 & 2$\times$1200 & 660 nm \\ 
         IC2554 & 001102 & 10 08 51.24 & -67 01 39.5 & 2$\times$420 & $\cdots$ & $\cdots$ & $\cdots$ \\
         IC5176 & 001106 & 22 14 52.33 & -66 51 31.7 & 2$\times$480 & 2$\times$300 & 2$\times$1800 & 660 nm \\ 
    ESO079-G007 & 001106 & 00 50 03.84 & -66 33 09.3 & 2$\times$480 & 2$\times$300 & 2$\times$1800 & 660 nm \\ 
    ESO035-G009 & 001106 & 07 28 40.95 & -75 03 14.9 & 2$\times$600 & 2$\times$600 & 2$\times$1200 & 659 nm \\
         IC2554 & 001106 & 10 08 51.24 & -67 01 39.5 & 2$\times$420 & 2$\times$300 & 2$\times$1200 & 659 nm \\

\enddata

\tablenotetext{a}{Right ascension from NED (2000.0)}
\tablenotetext{b}{Declination from NED (2000.0)}
\tablenotetext{c}{Exposure time}
\tablenotetext{d}{H$\alpha$ filter used during observation(s); the 657, 658, 659, 660, \& 661 nm filters are the 30 $\mbox{\AA}$ Rand filters; the 6563, 6600, 6618, \& 6653 $\mbox{\AA}$ filters are the NOAO 75 $\mbox{\AA}$ filters}
\label{tab1}
\end{deluxetable}

%% file: tab2.tex
\begin{deluxetable}{lll}
\tablecolumns{3}
\tablewidth{0pc}
\tablecaption{Standard Star Fields}
\tablehead{
\colhead{Observing run\tablenotemark{a}} & \colhead{Photometric Standards\tablenotemark{b}} & 
\colhead{Spectrophotmetric Standards\tablenotemark{c}}}

\startdata
CTIO, March 2-9, 2000 			& PG1525-071 & LTT 4364 \\
					& PG1525-071A & LTT 1788 \\
					& PG1525-071B & LTT 6248 \\
					& PG1525-071C & \\
					& PG1525-071D & \\
\hline
CTIO, Oct. 23-28, Nov. 2, 5, 6, 2000	& PG0231+051 & LTT 1020 \\
					& PG0231+051A & LTT 377 \\
					& PG0231+051B & \\
					& PG0231+051C & \\
					& PG0231+051D & \\
					& PG0231+051E & \\
					& GD 50 & \\
\enddata

\tablenotetext{a}{See section 2 for more detailed descriptions of these observing runs}
\tablenotetext{b}{The stars used as B and R standards as they are identified in \protect\cite{lan92}}
\tablenotetext{c}{The stars used as standards for the H$\alpha$ observations as they are identified in \protect\cite{sto83} and \protect\cite{bal84}}
\label{tab2}
\end{deluxetable}

%% file: tab3.tex
\begin{deluxetable}{lrrrr}
\tablecolumns{5}
\tablewidth{0pc}
\tablecaption{Photometric Calibration Coefficients}
\tablehead{
\colhead{Observing run\tablenotemark{a}} & \colhead{Filter\tablenotemark{b}} & \colhead{k\tablenotemark{c}} & 
\colhead{t\tablenotemark{d}} & \colhead{z\tablenotemark{e}}}

\startdata
CTIO, March 2-9, 2000 			& B & 0.28$\pm$0.031 & 0.090$\pm$0.010 & 1.80$\pm$0.041 \\
					& R & 0.13$\pm$0.018 & 0.050$\pm$0.0058 & 1.64$\pm$0.024 \\
					& 657 nm & 0.069$\pm$0.10 & $\cdots$ & 5.48$\pm$0.13 \\
					& 660 nm & 0.072$\pm$0.087 & $\cdots$ & 5.64$\pm$0.11 \\
					& 661 nm & 0.070$\pm$0.079 & $\cdots$ & 5.68$\pm$0.10 \\
					& 6600 $\mbox{\AA}$ & 0.135$\pm$0.093 & $\cdots$ & 4.65$\pm$0.12 \\
\hline
CTIO, Oct. 23-28, 2000 			& B & 0.25$\pm$0.010  & 0.061$\pm$0.0044  & 1.69$\pm$0.014  \\
					& R & 0.10$\pm$0.039  & -0.015$\pm$0.011  & 1.58$\pm$0.053  \\
					& 6563 $\mbox{\AA}$ & 0.071$\pm$0.051 & $\cdots$ & 4.64$\pm$0.061 \\
					& 6600 $\mbox{\AA}$ & 0.067$\pm$0.024 & $\cdots$ & 4.67$\pm$0.029 \\
\hline
CTIO, Nov. 2, 2000 			& B & 0.27$\pm$0.013  & 0.069$\pm$0.0034  & 1.68$\pm$0.018  \\
					& R & 0.10$\pm$0.0079  & -0.0024$\pm$0.0025  & 1.59$\pm$0.011  \\
\enddata

\tablenotetext{a}{Same as in Table 2}
\tablenotetext{b}{Same as in Table 1}
\tablenotetext{c}{The extinction coefficient through the filter in column (2) ($\pm$1$\sigma$)}
\tablenotetext{d}{The transformation coefficient through the filter in column (2) ($\pm$1$\sigma$)}
\tablenotetext{e}{The zero point through the filter in column (2) ($\pm$1$\sigma$)}
\label{tab3}
\end{deluxetable}

%% file: tab4.tex
\begin{deluxetable}{lrrrrrrr}
\tablecolumns{8}
\tablewidth{0pc}
\tablecaption{Isohpotal Data}
\tablehead{
\colhead{Galaxy} & \colhead{Type\tablenotemark{a}} & \colhead{r$_{25}$\tablenotemark{b}} & \colhead{m$_{B_{25}}$\tablenotemark{c}} & \colhead{m$_{R_{25}}$\tablenotemark{d}} & \colhead{E(B-V)\tablenotemark{e}} & \colhead{$i$\tablenotemark{f}} & \colhead{PA\tablenotemark{g}} \\
\colhead{} & \colhead{} & \colhead{($''$)} & \colhead{} & \colhead{} & \colhead{} & \colhead{($^{\circ}$)} & \colhead{($^{\circ}$)} }

\startdata
         ESO004-G017 &      Sm &     23 &  16.1 &  15.4 & 0.137 &     58 &    54 \\
         ESO013-G016 &      Sc &     51 &  13.5 &  12.5 & 0.072 &     48 &   -11 \\
         ESO017-G002 &      S0 &     27 &  14.6 &  13.0 & 0.248 &     18 &    -5 \\
         ESO019-G004 &      Sd &     19 &  15.4 &  14.3 & 0.125 &     70 &    31 \\
         ESO021-G003 &     Sbc &     39 &  13.8 &  12.6 & 0.242 &     51 &    60 \\
         ESO027-G001 &      Sc &     62 &  13.0 &   9.4 & 0.199 &     42 &   -43 \\
         ESO027-G021 &      Sa &     27 &  14.2 &  13.0 & 0.103 &     10 &   -90 \\
         ESO035-G009$\ast$$\dag$ &      Sm &     19 &  17.6 &  16.9 & 0.207 &     74 &    21 \\
         ESO035-G018$\ast$ &      Sc &     49 &  13.5 &  13.4 & 0.180 &     75 &   -42 \\
         ESO035-G020 &      Sm &     25 &  15.7 &  14.9 & 0.114 &     47 &    30 \\
         ESO035-G021 &      Sm &     19 &  16.3 &  14.9 & 0.183 &     69 &   128 \\
         ESO036-G006 &      Sd &     45 &  14.1 &  13.1 & 0.129 &     62 &   -18 \\
         ESO037-G004 &      Sm &     12 &  17.2 &  16.1 & 0.111 &     28 &   -10 \\
         ESO037-G010 &      Sc &     65 &  13.1 &  11.8 & 0.385 &     33 &   -16 \\
         ESO037-G015$\ast$ &     Sdm &     15 &  16.9 &  15.3 & 0.404 &     67 &   -67 \\
         ESO038-G011 &      Sb &     18 &  16.2 &  14.8 & 0.305 &     63 &    41 \\
         ESO054-G021 &     Scd &     90 &  12.7 &  13.1 & 0.051 &     61 &   -90 \\
         ESO059-G001 &      Sm &     51 &  14.2 &  13.1 & 0.146 &     31 &   -36 \\
         ESO060-G007$\ast$ &     Scd &     12 &  16.3 &  14.8 & 0.187 &     75 &    34 \\
         ESO060-G019 &      Sc &     67 &  12.1 &  11.3 & 0.100 &     61 &   -19 \\
        ESO060-IG003 &     Irr &     18 &  15.2 &  14.1 & 0.136 &     28 &    -6 \\
         ESO061-G017 &      Sd &     25 &  15.2 &  14.0 & 0.174 &     62 &   -54 \\
         ESO079-G005$\ast$ &     Scd &     42 &  13.9 &  11.0 & 0.020 &     59 &    -5 \\
         ESO079-G007$\ast$ &      Sc &     42 &  13.7 &  12.8 & 0.017 &     31 &     5 \\
         ESO080-G005$\ast$ &      Sm &     32 &  13.8 &  13.0 & 0.026 &     56 &    65 \\
         ESO084-G040 &      Sm &     28 &  15.1 &  14.4 & 0.035 &     46 &    57 \\
         ESO085-G014 &      Sm &     60 &  13.1 &  12.3 & 0.035 &     67 &    86 \\
         ESO085-G030 &     Irr &     31 &  13.9 &  12.9 & 0.030 &     63 &   -28 \\
         ESO085-G047 &     Irr &     32 &  15.1 &  14.3 & 0.025 &     53 &    29 \\
         ESO086-G060 &     Irr &     20 &  16.1 &  15.3 & 0.050 &     53 &   -63 \\
         ESO090-G004 &     IBm &     31 &  15.1 &  14.0 & 0.157 &     42 &   -88 \\
         ESO091-G007 &      Im &     44 &  13.6 &  12.5 & 0.193 &     60 &    -4 \\
         ESO092-G006 &      Sb &     47 &  13.1 &  11.8 & 0.227 &     55 &    35 \\
         ESO092-G021 &     Irr &     44 &  13.2 &  11.9 & 0.238 &     49 &    61 \\
         ESO104-G022 &      Im &     29 &  15.3 &  14.1 & 0.084 &     51 &   -72 \\
         ESO119-G016 &      Sm &     39 &  15.0 &  14.1 & 0.025 &     66 &    25 \\
         ESO140-G019 &     Irr &     18 &  16.6 &  15.9 & 0.080 &     47 &    34 \\
       HIPASS0635-70$\ast$ &      Sc &     12 &  17.7 &  16.7 & 0.069 &     46 &   -90 \\
       HIPASS0653-73 &      Sd &     12 &  17.9 &  16.8 & 0.133 &     62 &   -40 \\
       HIPASS1039-71 &     Irr &      9 &  17.8 &  16.5 & 0.149 &     33 &   -72 \\
              IC2150 &      Sc &     64 &  12.2 &  10.9 & 0.113 &     53 &    73 \\
              IC2554$\ast$ &  merger &     33 &  14.7 &  13.4 & 0.204 &     65 &     4 \\
              IC3104 &      Sm &     68 &  12.9 &  11.5 & 0.407 &     60 &    36 \\
              IC4662 &      Im &     81 &  11.2 &  10.4 & 0.070 &     48 &   -77 \\
              IC4710$\ast$ &      Sm &     80 &  12.9 &  11.7 & 0.089 &     15 &   -62 \\
              IC4870 &      Im &     38 &  13.9 &  15.0 & 0.113 &     58 &   -42 \\
              IC5028$\ast$ &      Sm &     28 &  15.6 &  14.9 & 0.048 &     53 &   -11 \\
              IC5176$\ast$ &     Sbc &     54 &  13.2 &  11.8 & 0.031 &     80 &    29 \\
                 M83 &     Sab &    279 &   8.5 &   7.5 & 0.066 &     13 &    85 \\
             NGC0406 &      Sc &     59 &  12.9 &  12.0 & 0.023 &     65 &   -19 \\
             NGC0802 &      Sa &     24 &  14.2 &  21.4 & 0.024 &     43 &   -20 \\
             NGC1313 &      Sc &    219 &  10.0 &   9.2 & 0.102 &     37 &     1 \\
             NGC1511 &      Sc &     68 &  12.0 &  10.9 & 0.061 &     68 &   -55 \\
             NGC1559 &     Scd &     90 &  10.9 &  10.0 & 0.030 &     53 &    65 \\
             NGC1809 &     Scd &     68 &  12.2 &  11.0 & 0.221 &     69 &   -37 \\
             NGC1892 &      Sc &     51 &  12.8 &  11.7 & 0.084 &     71 &    75 \\
             NGC2082 &      Sc &     51 &  12.7 &  11.6 & 0.058 &     24 &    51 \\
            NGC2397B &     Irr &     24 &  15.2 &  14.1 & 0.200 &     60 &    86 \\
             NGC2442$\ast$ &     Sbc &    130 &  11.6 &  10.0 & 0.203 &     21 &   -67 \\
            NGC2788B$\ast$ &     Sab &     27 &  14.9 &  13.7 & 0.111 &     64 &   -17 \\
             NGC2836 &     Sbc &     53 &  13.1 &  11.9 & 0.093 &     56 &   -69 \\
             NGC2915 &     Irr &     51 &  13.2 &  12.0 & 0.276 &     57 &   -53 \\
             NGC3059 &      Sc &     98 &  11.9 &  10.6 & 0.244 &     21 &    57 \\
            NGC3136A &      Im &     27 &  15.2 &  13.9 & 0.216 &     75 &    84 \\
             NGC5068$\ast$ &     SBd &    180 &  10.7 &   9.7 & 0.102 &     24 &   -76 \\
             NGC6300 &      Sb &    126 &  10.5 &   9.2 & 0.097 &     48 &   -65 \\
            NGC6438A &    Ring &     54 &  12.6 &  11.5 & 0.170 &     60 &    29 \\
             NGC7098 &      Sa &     91 &  12.4 &  11.1 & 0.087 &     51 &    70 \\
             NGC7661 &      Sc &     42 &  14.2 &  13.3 & 0.026 &     47 &    30 \\
\enddata

\tablenotetext{a}{Morphological type in the de Vaucouleurs classification system}
\tablenotetext{b}{Isophotal radius}
\tablenotetext{c}{B-band isophotal magnitude}
\tablenotetext{d}{R-band isophotal magnitude}
\tablenotetext{e}{Color excess}
\tablenotetext{f}{Inclination}
\tablenotetext{g}{Position angle measured from North through East}
\tablecomments{The images for any galaxy with a $\ast$ by its name were most likely taken through clouds (the same notation is used in Tables 5 and 6) and the values listed should be taken as estimates only.  For galaxies marked by a $\dag$, the data are for the 26 mag. arcsec.$^{-2}$ isophote.}
\label{tab4}
\end{deluxetable}

%% file: tab5.tex
\begin{deluxetable}{lrrrrrrrrr}
\tabletypesize{\small}
\tablecolumns{10}
\tablewidth{0pc}
\tablecaption{Surface Brightness Profile Data}
\tablehead{
\colhead{Galaxy} & \colhead{B$_{T}$\tablenotemark{a}} & \colhead{R$_{T}$\tablenotemark{b}} & \colhead{(B-R)$_{e}$\tablenotemark{c}} & \colhead{r$_{e}$\tablenotemark{d}} & \colhead{$\mu_{e,B}$\tablenotemark{e}} & \colhead{$\mu_{e,R}$\tablenotemark{f}} & \colhead{$h$\tablenotemark{g}} & \colhead{$\mu_{o,B}$\tablenotemark{h}} & \colhead{$\mu_{o,R}$\tablenotemark{i}} \\
\colhead{} & \colhead{} & \colhead{} & \colhead{} & \colhead{($''$)} & \colhead{(mag/$\Box ''$)} & \colhead{(mag/$\Box ''$)} & \colhead{($''$)} & \colhead{(mag/$\Box ''$)} & \colhead{(mag/$\Box ''$)} }

\startdata   
         ESO004-G017 &  14.5 &  13.9 &  0.56 &   31.1 &  24.7 &  24.1 &   19.4 &  22.9 &  22.4 \\
         ESO013-G016 &  13.0 &  12.2 &  0.86 &   25.9 &  22.8 &  21.9 &   11.0 &  19.7 &  18.8 \\
         ESO017-G002 &  13.3 &  12.2 &  1.13 &    9.3 &  21.1 &  20.1 &    9.1 &  20.6 &  19.5 \\
         ESO019-G004 &  14.8 &  13.9 &  0.80 &   12.4 &  23.1 &  22.4 &    5.5 &  20.7 &  19.9 \\
         ESO021-G003 &  12.6 &  11.8 &  0.82 &   22.2 &  22.4 &  21.7 &   11.5 &  20.3 &  19.5 \\
         ESO027-G001 &  11.8 &  10.9 &  0.99 &   37.0 &  22.4 &  21.6 &   13.2 &  18.6 &  17.9 \\
         ESO027-G021 &  13.7 &  12.7 &  1.03 &    7.1 &  21.0 &  19.9 &    6.6 &  20.0 &  19.2 \\
         ESO035-G009$\ast$ &  15.7 &  13.7 &  1.99 &   24.2 &  25.6 &  23.7 &   18.4 &  24.1 &  22.2 \\
         ESO035-G018$\ast$ &  12.6 &  11.6 &  1.06 &   21.8 &  22.4 &  21.5 &   14.3 &  20.6 &  19.9 \\
         ESO035-G020 &  14.4 &  13.6 &  0.72 &   25.2 &  24.4 &  23.9 &   16.4 &  22.8 &  22.1 \\
         ESO035-G021 &  12.3 &  12.0 &  0.41 &   25.4 &  24.3 &  23.9 &   25.9 &  23.3 &  22.8 \\
         ESO036-G006 &  13.1 &  12.3 &  0.72 &   35.5 &  24.1 &  23.2 &   23.9 &  22.4 &  21.6 \\
         ESO037-G004 &  14.0 &  13.3 &  0.67 &   65.2 &  25.8 &  25.2 &   40.5 &  24.1 &  23.4 \\
         ESO037-G010 &  11.0 &  10.4 &  0.59 &   36.1 &  22.1 &  21.3 &   32.8 &  21.1 &  20.3 \\
         ESO037-G015$\ast$ &  14.4 &  13.6 &  0.95 &   15.8 &  23.1 &  22.6 &   10.6 &  21.7 &  21.0 \\
         ESO038-G011 &  13.9 &  12.7 &  1.04 &   15.0 &  23.0 &  22.3 &    7.1 &  20.8 &  20.1 \\
         ESO054-G021 &  12.2 &  11.3 &  0.85 &   54.8 &  23.6 &  22.8 &   31.0 &  21.6 &  20.8 \\
         ESO059-G001 &  12.6 &  11.9 &  0.80 &   52.8 &  24.3 &  23.7 &   27.2 &  22.3 &  21.6 \\
         ESO060-G007$\ast$ &  15.5 &  14.5 &  1.16 &   13.7 &  24.3 &  23.5 &   11.0 &  23.0 &  22.1 \\
         ESO060-G019 &  11.6 &  10.9 &  0.63 &   34.8 &  22.9 &  22.1 &   19.0 &  20.6 &  20.0 \\
        ESO060-IG003 &  14.5 &  13.6 &  0.70 &    6.7 &  21.9 &  21.1 &    4.9 &  20.4 &  19.4 \\
         ESO061-G017 &  14.3 &  13.3 &  0.93 &   16.0 &  23.0 &  22.1 &    8.9 &  21.1 &  20.2 \\
         ESO079-G005$\ast$ &  13.7 &  12.8 &  0.92 &   20.8 &  23.0 &  22.2 &   12.3 &  21.2 &  20.4 \\
         ESO079-G007$\ast$ &  13.5 &  12.6 &  0.83 &   20.8 &  22.8 &  22.0 &   12.1 &  20.9 &  19.9 \\
         ESO080-G005$\ast$ &  13.6 &  12.8 &  0.84 &   18.0 &  23.1 &  22.4 &    8.6 &  20.7 &  20.0 \\
         ESO084-G040 &  14.7 &  14.1 &  0.64 &   15.3 &  23.5 &  22.9 &    9.6 &  21.7 &  21.2 \\
         ESO085-G014 &  12.8 &  12.1 &  0.73 &   30.7 &  23.0 &  22.3 &   14.9 &  20.5 &  19.9 \\
         ESO085-G030 &  13.7 &  12.8 &  0.93 &    9.2 &  21.2 &  20.4 &    7.1 &  20.3 &  19.4 \\
         ESO085-G047 &  13.9 &  13.2 &  0.77 &   45.8 &  25.2 &  24.6 &   31.9 &  23.4 &  22.7 \\
         ESO086-G060 &  15.6 &  15.0 &  0.65 &   13.2 &  24.0 &  23.4 &    8.0 &  22.2 &  21.7 \\
         ESO090-G004 &  14.0 &  13.2 &  0.74 &   21.0 &  23.3 &  22.6 &   13.2 &  21.7 &  21.1 \\
         ESO091-G007 &  12.6 &  11.9 &  0.77 &   24.6 &  22.4 &  21.8 &   12.5 &  20.3 &  19.6 \\
         ESO092-G006 &  10.9 &  10.6 &  0.64 &   48.3 &  23.9 &  23.4 &   13.4 &  19.9 &  19.4 \\
         ESO092-G021 &  11.4 &  10.6 &  0.92 &   42.1 &  23.6 &  23.0 &   33.6 &  22.0 &  21.4 \\
         ESO104-G022 &  13.7 &  12.8 &  1.03 &   46.0 &  25.1 &  24.3 &   35.9 &  23.7 &  22.9 \\
         ESO119-G016 &  14.2 &  13.5 &  0.77 &   35.6 &  24.7 &  24.0 &   21.9 &  23.0 &  22.3 \\
         ESO140-G019 &  15.2 &  14.6 &  0.56 &   22.9 &  24.8 &  24.2 &   14.9 &  23.1 &  22.6 \\
       HIPASS0635-70$\ast$ &  16.7 &  15.8 &  0.92 &   11.2 &  24.6 &  23.9 &    7.0 &  23.0 &  22.0 \\
       HIPASS0653-73 &  16.1 &  15.2 &  0.82 &   17.6 &  25.0 &  24.2 &   10.9 &  23.3 &  22.5 \\
       HIPASS1039-71 &  16.9 &  15.8 &  1.16 &    7.2 &  24.0 &  23.2 &    4.2 &  22.2 &  21.3 \\
              IC2150 &  11.7 &  10.6 &  1.27 &   30.8 &  21.7 &  20.7 &   10.5 &  17.8 &  16.7 \\
              IC2554$\ast$ &  13.8 &  12.8 &  1.08 &   19.2 &  22.9 &  22.0 &   11.0 &  21.1 &  20.2 \\
              IC3104 &  10.8 &  10.1 &  0.55 &   31.1 &  21.7 &  20.9 &   28.6 &  20.6 &  19.7 \\
              IC4662 &  10.8 &  10.0 &  0.62 &   37.3 &  22.1 &  21.3 &   21.7 &  20.5 &  19.5 \\
              IC4710$\ast$ &  12.1 &  11.1 &  1.07 &   60.3 &  23.6 &  22.9 &   40.2 &  22.3 &  21.5 \\
              IC4870 &  13.1 &  14.4 & -1.15 &   20.8 &  23.0 &  24.7 &   13.2 &  21.3 &  22.9 \\
              IC5028$\ast$ &  14.9 &  14.3 &  0.72 &   21.2 &  24.2 &  23.6 &   12.8 &  22.5 &  21.9 \\
              IC5176$\ast$ &  13.0 &  11.7 &  1.50 &   23.0 &  22.7 &  21.5 &   13.9 &  20.7 &  19.7 \\
                 M83 &   8.2 &   7.3 &  1.06 &  126.8 &  21.4 &  20.5 &   21.9 &  10.8 &   9.7 \\
             NGC0406 &  12.7 &  11.8 &  0.95 &   25.3 &  22.6 &  21.7 &   19.3 &  21.5 &  20.6 \\
             NGC0802 &  14.0 &  13.2 &  0.67 &    5.5 &  20.7 &  19.9 &    8.1 &  21.6 &  20.6 \\
             NGC1313 &   9.5 &   8.8 &  0.63 &   94.6 &  22.5 &  21.7 &   59.3 &  20.6 &  19.9 \\
             NGC1511 &  11.7 &  10.6 &  0.99 &   20.9 &  21.2 &  20.1 &   17.7 &  20.6 &  19.4 \\
             NGC1559 &  10.7 &   9.8 &  0.85 &   33.0 &  21.0 &  20.1 &   18.9 &  19.6 &  18.7 \\
             NGC1809 &  10.7 &   9.9 &  0.76 &   60.5 &  23.2 &  22.8 &   30.0 &  21.2 &  20.8 \\
             NGC1892 &  12.3 &  11.4 &  0.90 &   21.8 &  21.7 &  20.9 &   10.3 &  19.3 &  18.4 \\
             NGC2082 &  12.4 &  11.4 &  1.03 &   22.0 &  21.5 &  20.6 &   10.7 &  19.7 &  18.5 \\
            NGC2397B &  14.2 &  13.3 &  0.80 &   14.6 &  22.9 &  22.1 &    9.1 &  21.2 &  20.3 \\
             NGC2442$\ast$ &  10.3 &   9.1 &  1.29 &  104.3 &  23.2 &  22.1 &   65.9 &  21.6 &  20.5 \\
            NGC2788B$\ast$ &  14.3 &  13.2 &  0.90 &   13.7 &  22.8 &  21.8 &    9.6 &  21.6 &  20.3 \\
             NGC2836 &  12.5 &  11.4 &  1.09 &   26.8 &  22.7 &  21.6 &   16.6 &  21.0 &  19.9 \\
             NGC2915 &  11.7 &  11.0 &  0.58 &   18.7 &  21.3 &  20.4 &   17.5 &  20.6 &  19.7 \\
             NGC3059 &  10.2 &   9.5 &  0.88 &   74.4 &  23.0 &  22.2 &   38.7 &  20.9 &  20.2 \\
            NGC3136A &  14.1 &  13.3 &  0.79 &   19.9 &  23.5 &  22.7 &   11.9 &  21.7 &  20.9 \\
             NGC5068$\ast$ &  10.1 &   9.3 &  0.94 &   97.8 &  22.8 &  22.0 &   46.1 &  20.3 &  19.5 \\
             NGC6300 &  10.0 &   8.8 &  1.12 &   56.2 &  21.6 &  20.5 &   31.6 &  20.1 &  18.8 \\
            NGC6438A &  10.7 &   9.9 &  0.89 &   59.0 &  23.9 &  23.3 &   65.0 &  22.7 &  22.1 \\
             NGC7098 &  11.6 &  10.2 &  1.49 &   71.6 &  24.2 &  23.0 &   62.2 &  23.0 &  21.8 \\
             NGC7661 &  13.9 &  13.1 &  0.98 &   24.1 &  23.4 &  22.6 &    7.3 &  18.7 &  17.9 \\
\enddata

\tablenotetext{a}{Total extrapolated B-band magnitude}
\tablenotetext{b}{Total extrapolated R-band magnitude} 
\tablenotetext{c}{Color index within the effective radius}
\tablenotetext{d}{Effective radius}
\tablenotetext{e}{B-band surface brightness at the effective radius}
\tablenotetext{f}{R-band surface brightness at the effective radius}
\tablenotetext{g}{Disk scale length}
\tablenotetext{h}{Extrapolated B-band disk central surface brightness}
\tablenotetext{i}{Extrapolated R-band disk central surface brightness}
\tablecomments{All colors and central surface brightnesses were corrected for Galactic extinction using E(B-V) values quoted in the NASA/IPAC Extragalactic Database (NED) and E(B-V) to A$_{\nu}$ conversion factors provided by Schlegel et al. (1998).}
\label{tab5}
\end{deluxetable}

%% file: tab6.tex
\begin{deluxetable}{lrrrrrrrrr}
\tabletypesize{\scriptsize}
\tablecolumns{10}
\tablewidth{0pc}
\tablecaption{Emission Line Data}
\tablehead{
\colhead{Galaxy} & \colhead{log $F_{H\alpha+[NII]}$\tablenotemark{a}} & \colhead{log $F_{H\alpha}$\tablenotemark{b}} & \colhead{$W_{H\alpha+[NII]}$\tablenotemark{c}} & \colhead{$W_{H\alpha}$\tablenotemark{d}} & 
\colhead{$S_{HI}$\tablenotemark{e}} & \colhead{$\frac{M_{HI}}{L_{B}}$\tablenotemark{f}} &  \colhead{$\frac{M_{HI}}{L_{R}}$\tablenotemark{g}} & \colhead{$V_{R}$\tablenotemark{h}} & \colhead{$V_{LG}$\tablenotemark{i}} \\
\colhead{} & \colhead{(ergs/s/cm$^{2}$)} & \colhead{(ergs/s/cm$^{2}$)} & \colhead{($\mbox{\AA}$)} & \colhead{($\mbox{\AA}$)} & \colhead{(Jy km/s)} & \colhead{(M$_{\odot}$/L$_{\odot}$)} & \colhead{(M$_{\odot}$/L$_{\odot}$)} & \colhead{(km/s)} & \colhead{(km/s)} }

\startdata
         ESO004-G017 &   -13.23 &   -12.92 &     30.2 &     29.2 &   4.2 & 0.39 & 0.70 & 1754 & 1750 \\
         ESO013-G016$\star$ &   -12.13 &   -11.83 &     21.4 &     18.0 &   2.3 & 0.06 & 0.08 & 1753 & 1737 \\
         ESO017-G002$\star$ &   -12.72 &   -12.26 &      8.9 &      7.5 &   4.9 & 0.16 & 0.17 & 1620 & 1607 \\
         ESO019-G004 &   -13.11 &   -12.79 &     14.5 &     14.4 &   2.3 & 0.28 & 0.38 & 2005 & 2013 \\
         ESO021-G003 &   -12.02 &   -11.41 &     14.1 &     13.4 &   5.5 & 0.10 & 0.14 & 2283 & 2287 \\
         ESO027-G001$\star$ &   -10.10 &    -9.43 &    131.8 &    111.9 &  29.1 & 0.24 & 0.30 & 2548 & 2543 \\
         ESO027-G021$\star$ &   -12.34 &   -11.96 &     21.4 &     19.1 &  10.2 & 0.47 & 0.55 & 2443 & 2433 \\
         ESO035-G009$\ast$$\star$ & $\cdots$ & $\cdots$ & $\cdots$ & $\cdots$ &   5.2 & 1.49 & 0.71 & 1122 & 1096 \\
         ESO035-G018$\ast$$\star$ &   -12.63 &   -12.13 &     14.5 &     13.6 &  15.6 & 0.25 & 0.32 & 1750 & 1745 \\
         ESO035-G020 &   -12.46 &   -12.13 &    112.2 &    107.6 &  10.8 & 0.91 & 1.37 & 1742 & 2211 \\
         ESO035-G021 &   -13.07 &   -12.73 &     26.3 &     20.8 &   3.8 & 0.05 & 0.11 & 1235 & 1215 \\
         ESO036-G006 &   -12.50 &   -12.21 &     20.9 &     17.9 &  24.0 & 0.64 & 0.89 & 1132 & 1112 \\
         ESO037-G004 &   -13.45 &   -13.22 &     37.2 &     32.0 &   4.0 & 0.24 & 0.39 & 1274 & 1263 \\
         ESO037-G010 &   -11.91 &   -11.06 &     21.4 &     19.6 &  16.5 & 0.07 & 0.11 & 1773 & 1781 \\
         ESO037-G015$\ast$ &   -13.23 &   -12.66 &     30.9 &     29.1 &   3.9 & 0.35 & 0.50 & 1610 & 1612 \\
         ESO038-G011 &   -13.20 &   -12.64 &     20.0 &     19.4 &   4.5 & 0.25 & 0.26 & 1879 & 1887 \\
         ESO054-G021$\star$ &   -12.07 &   -11.78 &     60.3 &     48.6 &  38.8 & 0.44 & 0.61 & 1425 & 1379 \\
         ESO059-G001 &   -12.48 &   -12.19 &     20.4 &     20.1 &  14.2 & 0.24 & 0.36 &  530 &  448 \\
         ESO060-G007$\ast$ &   -13.49 &   -13.19 &     10.2 &      9.6 &   0.1 & 0.03 & 0.03 & 1506 & 1502 \\
         ESO060-G019 &   -11.66 &   -11.27 &     21.4 &     17.3 &  44.4 & 0.28 & 0.47 & 1429 & 1428 \\
        ESO060-IG003 &   -12.94 &   -12.67 &     17.8 &     16.1 &   2.5 & 0.24 & 0.33 & 1415 & 1401 \\
         ESO061-G017 &   -13.10 &   -12.72 &      9.1 &      8.8 &   5.4 & 0.43 & 0.52 & 1748 & 1765 \\
         ESO079-G005$\ast$$\star$ &   -10.50 &   -10.26 &    263.0 &    250.0 &  21.1 & 0.97 & 1.31 & 1701 & 1687 \\
         ESO079-G007$\ast$$\star$ &   -12.13 &   -11.91 &     34.7 &     30.5 &   7.3 & 0.28 & 0.35 & 1476 & 1629 \\
         ESO080-G005$\ast$$\star$ &   -12.42 &   -12.21 &     20.0 &     18.2 &   7.9 & 0.35 & 0.49 & 1549 & 1486 \\
         ESO084-G040 &   -13.11 &   -12.99 &     13.5 &     11.9 &   0.5 & 0.06 & 0.10 & 1236 & 1163 \\
         ESO085-G014 &   -11.79 &   -11.56 &     45.7 &     39.2 &  21.0 & 0.42 & 0.67 & 1401 & 1327 \\
         ESO085-G030$\star$ &   -12.09 &   -11.91 &     34.7 &     30.3 &  15.7 & 0.72 & 0.93 & 1289 & 1223 \\
         ESO085-G047 &   -12.81 &   -12.63 &     25.1 &     22.7 &   9.8 & 0.55 & 0.82 & 1458 & 1385 \\
         ESO086-G060 &   -13.39 &   -13.22 &     18.6 &     18.0 &   3.9 & 1.06 & 1.82 & 1636 & 1585 \\
         ESO090-G004 &   -12.57 &   -12.17 &     38.0 &     37.2 &   4.1 & 0.25 & 0.37 & 1924 & 2083 \\
         ESO091-G007 &   -12.49 &   -11.96 &      3.3 &      3.0 &   6.4 & 0.11 & 0.16 & 2130 & 2174 \\
         ESO092-G006 &   -11.80 &   -11.15 &     12.0 &     10.9 &  15.0 & 0.05 & 0.12 & 2166 & 1739 \\
         ESO092-G021 &   -12.11 &   -11.45 &      7.4 &      6.1 &  28.3 & 0.16 & 0.22 & 2008 & 2040 \\
         ESO104-G022 &   -13.00 &   -12.75 &     17.0 &     17.0 &  15.2 & 0.70 & 0.88 &  794 &  804 \\
         ESO119-G016$\star$ &   -12.64 &   -12.48 &     32.4 &     31.7 &  11.5 & 0.87 & 1.29 &  961 &  882 \\
         ESO140-G019 &   -12.80 &   -12.61 &    134.9 &    134.9 &   4.1 & 0.73 & 1.32 &  959 &  975 \\
       HIPASS0635-70$\ast$ &   -14.50 &   -14.34 &      3.0 &      2.9 &   4.3 & 3.06 & 3.96 & 1611 & 1582 \\
       HIPASS0653-73 &   -14.03 &   -13.82 &     14.8 &     13.8 &   3.9 & 1.59 & 2.25 & 1202 & 1171 \\
       HIPASS1039-71 &   -14.70 &   -14.46 &      2.9 &      2.8 &   2.4 & 2.06 & 2.24 & 1537 & 1539 \\
              IC2150$\star$ &   -11.35 &   -10.89 &     31.6 &     24.2 &  17.4 & 0.13 & 0.13 & 1724 & 1709 \\
              IC2554$\ast$$\star$ &   -12.20 &   -11.79 &     60.3 &     58.6 &  14.0 & 0.74 & 0.88 & 1386 & 1388 \\
              IC3104 &   -12.29 &   -11.67 &      7.6 &      7.3 &   6.5 & 0.02 & 0.03 &  432 &  415 \\
              IC4662 &   -10.85 &   -10.58 &     70.8 &     67.3 &  99.0 & 0.31 & 0.46 &  304 &  391 \\
              IC4710$\ast$ &   -11.82 &   -11.49 &     33.1 &     33.0 &  25.9 & 0.27 & 0.31 &  737 &  744 \\
              IC4870$\star$ &   -12.22 &   -12.03 &     12.0 &     11.1 &  18.8 & 0.50 & 5.04 &  876 &  884 \\
              IC5028$\ast$$\star$ &   -12.82 &   -12.63 &     49.0 &     46.7 &   4.8 & 0.70 & 1.12 & 1620 & 1652 \\
              IC5176$\ast$$\star$ &   -12.12 &   -11.77 &     13.2 &     12.3 &  42.1 & 0.98 & 0.90 & 1723 & 1745 \\
                 M83 &   -10.01 &    -9.27 &     35.5 &     32.8 & 417.3 & 0.12 & 0.15 &  516 &  751 \\
             NGC0406$\star$ &   -11.66 &   -11.41 &     46.8 &     38.5 &  34.4 & 0.61 & 0.79 & 1504 & 1473 \\
             NGC0802$\star$ &   -13.65 &   -13.48 &   1621.8 &   1422.3 &   2.5 & 0.15 & 0.21 & 1477 & 1428 \\
             NGC1313 &   -10.59 &   -10.21 &     34.7 &     33.3 & 219.5 & 0.21 & 0.34 &  457 &  350 \\
             NGC1511$\star$ &   -11.17 &   -10.83 &     46.8 &     36.4 &  59.9 & 0.42 & 0.48 & 1330 & 1271 \\
             NGC1559$\star$ &   -10.72 &   -10.36 &     61.7 &     45.7 &  33.3 & 0.09 & 0.13 & 1301 & 1224 \\
             NGC1809$\star$ &   -11.82 &   -11.27 &     11.2 &      8.4 &  24.2 & 0.07 & 0.11 & 1296 & 1246 \\
             NGC1892 &   -11.69 &   -11.39 &     35.5 &     28.8 &  37.8 & 0.49 & 0.65 & 1352 & 1288 \\
             NGC2082$\star$ &   -11.74 &   -11.48 &     24.0 &     19.8 &   6.7 & 0.09 & 0.11 & 1179 & 1124 \\
            NGC2397B &   -12.96 &   -12.62 &     14.8 &     13.2 &   4.0 & 0.30 & 0.39 & 1401 & 1383 \\
             NGC2442$\ast$ &   -11.42 &   -10.72 &     12.0 &      8.8 &  51.3 & 0.10 & 0.10 & 1469 & 1453 \\
            NGC2788B$\ast$ &   -12.88 &   -12.62 &     12.9 &     11.5 &   4.9 & 0.39 & 0.41 & 1413 & 1407 \\
             NGC2836 &   -11.68 &   -11.26 &     39.8 &     36.6 &   9.4 & 0.14 & 0.15 & 1678 & 1688 \\
             NGC2915 &   -11.95 &   -11.51 &     25.1 &     24.5 &  83.1 & 0.63 & 0.94 &  452 &  402 \\
             NGC3059 &   -11.20 &   -10.59 &     35.5 &     25.5 &  56.7 & 0.11 & 0.17 & 1247 & 1234 \\
            NGC3136A &   -12.41 &   -11.99 &     29.5 &     26.7 &   3.4 & 0.23 & 0.32 & 1995 & 2025 \\
             NGC5068$\ast$ &   -10.91 &   -10.37 &     33.1 &     32.5 & 105.8 & 0.18 & 0.25 &  675 &  882 \\
             NGC6300 &   -11.01 &   -10.50 &     16.6 &     11.3 &  44.9 & 0.07 & 0.07 & 1103 & 1120 \\
            NGC6438A &   -11.82 &   -10.95 &      6.2 &      6.2 &  15.9 & 0.05 & 0.07 & 2550 & 2548 \\
             NGC7098$\star$ &   -11.81 &   -11.21 &     14.8 &     11.7 &  21.0 & 0.14 & 0.12 & 2339 & 2352 \\
             NGC7661$\star$ &   -12.34 &   -12.11 &     28.2 &     24.8 &  10.2 & 0.59 & 0.79 & 2022 & 2026 \\
\enddata

\tablenotetext{a}{Observed H$\alpha+[$NII$]$ flux}
\tablenotetext{b}{H$\alpha$ flux (corrected for $[$NII$]$ emission and Galacitc and internal extinction)}
\tablenotetext{c}{Observed H$\alpha+[$NII$]$ equivalent width}
\tablenotetext{d}{H$\alpha$ equivalent width (corrected for $[$NII$]$ emission)}
\tablenotetext{e}{Integrated 21 cm flux}
\tablenotetext{f}{B-band gas-to-light ratio (corrected for Galactic extinction)}
\tablenotetext{g}{R-band gas-to-light ratio (corrected for Galactic extinction)}
\tablenotetext{h}{Measured radial velocity}
\tablenotetext{i}{Radial velocity corrected for Virgo-centric infall relative to the rest frame of the Local Group}
\tablecomments{The H$\alpha$ flux for any galaxy with a $\star$ by its name was calibrated using the R-band flux.}
\label{tab6}
\end{deluxetable}

%% file: tab7.tex
\begin{deluxetable}{lrr}
\tablecolumns{3}
\tablewidth{0pc}
\tablecaption{K-S Test Results}
\tablehead{
\colhead{Property} & \colhead{log $P_{T}$\tablenotemark{a}} & \colhead{log $P_{S}$\tablenotemark{b}}}

\startdata
         $\mu_{e,B}$ & -4.6  & -3.5  \\
         $\mu_{e,R}$ & -8.0  & -5.1  \\
         $\mu_{o,B}$ & -1.8  & -0.38 \\
         $\mu_{o,R}$ & -0.31 & -0.48 \\
         (B-R)$_{e}$ & -10.1 & -5.5  \\
         log $r_{e}$ & -0.41 & -0.29 \\
             log $h$ & -3.1  & -1.6  \\
             $M_{B}$ & -1.8  & -0.54 \\
             $M_{R}$ & -2.3  & -1.4  \\
             log SFR & -2.0  & -2.1  \\
        log M$_{HI}$ & -1.5  & -2.2  \\
log M$_{HI}$/L$_{B}$ & -2.7  & -4.7  \\
log M$_{HI}$/L$_{R}$ & -1.5  & -3.0  \\
         $f_{g}$ & -0.91 & -1.8  \\
\enddata

\tablenotetext{a}{K-S test P value using all galaxy types}
\tablenotetext{b}{K-S test P value using only spiral galaxies}
\label{tab7}
\end{deluxetable}